\documentclass[reprint,
superscriptaddress,
bibnotes,
amsmath,amssymb,
aps,
floatfix]{revtex4-2}

\pdfoutput=1
\usepackage{graphicx}% Include figure files
\usepackage{dcolumn}% Align table columns on decimal point
\usepackage{bm}% bold math
\usepackage[colorlinks, citecolor={blue}, urlcolor={blue}, linkcolor={blue}]{hyperref}% add hypertext capabilities

\usepackage{multirow}
\usepackage{makecell}
\usepackage{wrapfig}
\usepackage{braket}
\usepackage{array}
\newcolumntype{L}{>{\centering\arraybackslash}m{.12\textwidth}}
\newcolumntype{M}{>{\centering\arraybackslash}m{.2\textwidth}}
\newcolumntype{N}{>{\centering\arraybackslash}m{.08\textwidth}}
\newcolumntype{P}{>{\centering\arraybackslash}m{.17\textwidth}}
\newcolumntype{Q}{>{\arraybackslash}p{5cm}}

\newcommand{\exb}{\vec{E}\times\vec{B}}

\begin{document}

\newcommand{\NIST}{National Institute of Standards and Technology, Boulder, Colorado, 80305}
\newcommand{\CU}{Department of Physics, University of Colorado, Boulder, Colorado 80309, USA}
\title{Localized control of large ion crystals in a Penning trap using a spatial light modulator}

\author{Allison L. Carter}
\email{allison.carter@nist.gov}
\affiliation{\NIST}
\author{Jennifer F. Lilieholm}
\affiliation{\NIST}
\author{Bryce B. Bullock}
\affiliation{\NIST}
\affiliation{\CU}
\author{Kurt Thompson}
\affiliation{\NIST}
\affiliation{\CU}
\author{Diep Nguyen}
\affiliation{\NIST}
\affiliation{\CU}
\author{John J. Bollinger}
\affiliation{\NIST}

\begin{abstract}
Penning ion traps as quantum platforms have primarily utilized global control and symmetric Dicke states for quantum simulation and sensing experiments. The introduction of local control greatly increases the power of the platform as a quantum simulator but is technically challenging due to the rapid rotation of the ion crystals. Here we use an ultraviolet-compatible spatial light modulator (SLM) to imprint programmable AC Stark shift patterns with different azimuthal symmetries and gradients that co-rotate with the ion crystals, demonstrating localized coherent control of single plane crystals with greater than 100 ions. Comparisons of the measured ion qubit populations with calculations from independent measurements of the applied AC Stark shift patterns show good agreement, validating the technique and providing a path, with a higher format SLM, for parallelizable, coherent individual ion addressing in Penning traps.
\end{abstract}

\maketitle

\textit{Introduction}--Penning traps, which employ static electric and magnetic fields to trap atomic ions, have shown great promise as a platform for quantum sensing~\cite{gilmore2021} and simulation~\cite{bohnet2016,bullock2026,jee2026} because of their utility in forming, cooling, and controlling large ion crystals~\cite{morigi2026}. Various elements of quantum control, including global qubit rotations and multi-qubit entangling operations obtained by uniformly coupling every ion qubit with the center-of-mass (COM) motional mode, have been demonstrated with up to hundreds of ions in two-dimensional crystals~\cite{gilmore2021,bohnet2016,bullock2026,britton2012}. These control elements enable quantum information processing within the Dicke manifold of symmetric spin states. Local control greatly expands the types of quantum simulations that can be performed and the Hilbert space that can be accessed. It can be used for preparing spatially non-uniform spin states as well as engineering spin-spin interactions, which can enable the simulation of a wider variety of systems. Reference~\onlinecite{shankar2022} shows, for example, how controlling a non-uniform coupling of the spins with the ion crystal center-of-mass mode can engineer interactions that simulate dynamical phases of chiral $p+ip$ superconductors.

Radio frequency (RF) Paul ion traps and neutral atom platforms, where the trapping potentials result in approximately stationary confinement, commonly use tightly focused laser beams for individual addressing and control~\cite{kranzl2022,debnath2016,nagerl1999,lee2016,wang2015,isenhower2010,schunke2015,kalathur2024,shih2021,mazzanti2024,chen2026} and as optical tweezers~\cite{kaufman2021,saffman2016,browaeys2020,kaufman2012,schwerdt2026,mazzanti2024,chen2026,schneider2010}, which are often generated with spatial light modulators (SLMs)~\cite{bergamini2004,kaufman2021}. The wavelengths for working with most of the commonly trapped ions are in the ultraviolet (UV), which significantly limits the available options for generating many beams for tweezers or addressing. We trap $^9$Be$^+$ ions, which have a transition wavelength of 313 nm, below the wavelength range where SLMs have been used for control in atomic platforms \cite{ammenwerth2025}. Acousto-optic modulators (AOMs) or acousto-optic deflectors (AODs) are frequently used when working with UV or blue lasers~\cite{kranzl2022,debnath2016,nagerl1999,lee2016}. The capabilities of these devices are more constrained than those of SLMs and generally require changes in the frequency to achieve a different position~\cite{pogorelov2021}.

Any individual addressing approach will be more challenging in Penning traps due to the rapid crystal rotation (180~kHz in our system). As in RF traps, tightly focused beams can still be used, either to drive qubit rotations with an AC Stark shift (ACSS) \cite{mcmahon2024,makadia2026} or to selectively incoherently repump some ion qubits \cite{jee2026}. References~\onlinecite{mcmahon2024,makadia2026} used a single, tightly focused beam with intensity and pulse duration control to drive a programmable rotation of a targeted ion qubit when it passed through the beam as the crystal rotated. While the beam was held at a fixed radius in Ref.~\onlinecite{mcmahon2024}, sequential addressing of every qubit in the crystal was achieved in Ref.~\onlinecite{makadia2026} by steering the beam with an AOM.

\begin{figure}[t]
    \centering
    \includegraphics[width=\columnwidth]{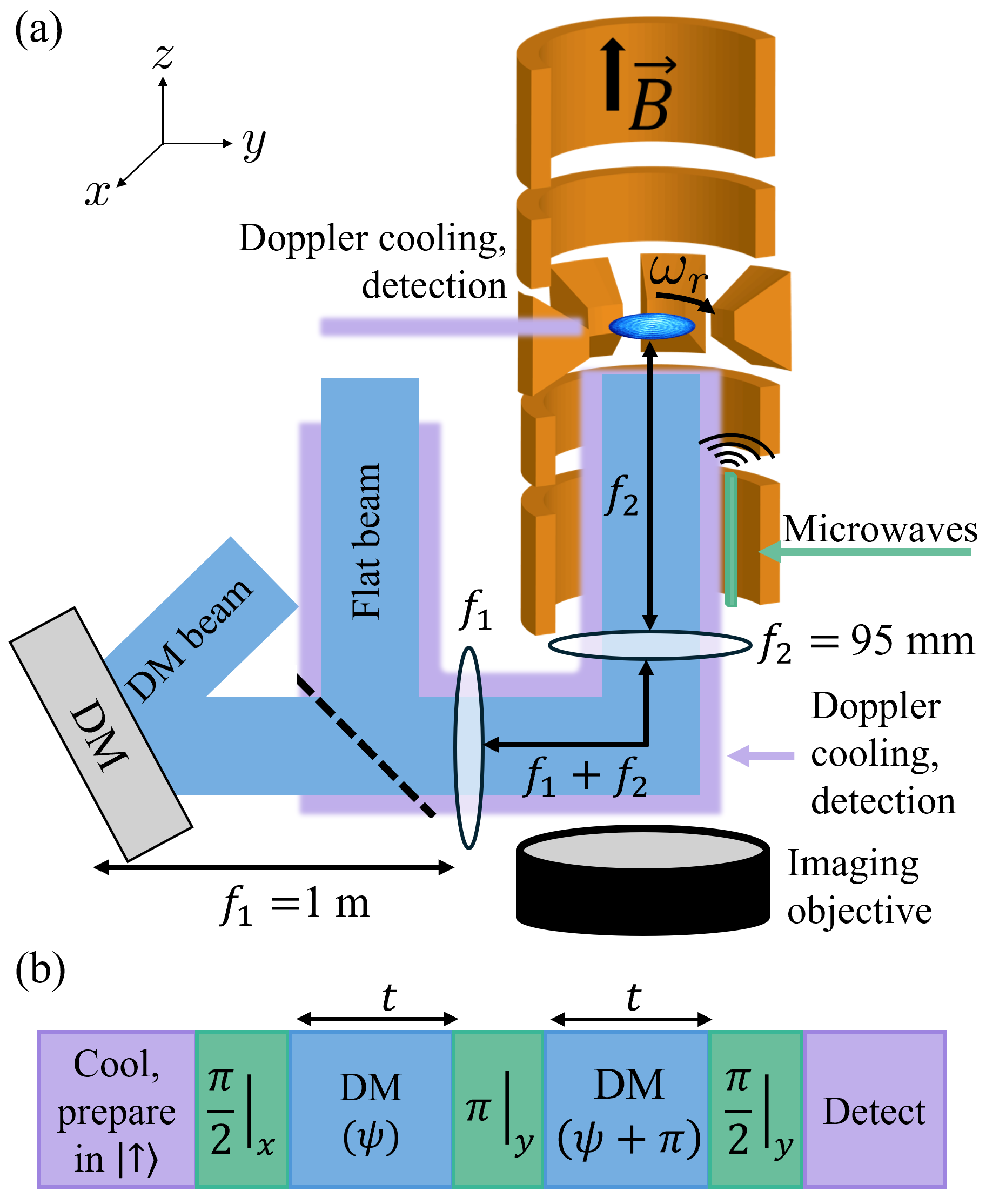}
    \caption{(a) Penning trap electrode and laser beam schematic (not to scale). The copper-colored segments are a cross-section of the trap electrodes. Doppler cooling and detection beams are sent parallel and perpendicular to $\vec{B}$. The deformable mirror (DM) beam is combined with the flat and parallel Doppler cooling beams with a 50/50 beam splitter. All three beams co-propagate along $\vec{B}$ with lenses with focal lengths $f_1$ and $f_2$ that image the DM surface. A microwave waveguide for global coherent rotations is located below the ions. Ion fluorescence is collected parallel to $\vec{B}$ from below the trap. (b) Experimental pulse sequence. The DM Hamiltonian is applied for an arm time $t$ in a spin-echo sequence, and spin precession is mapped into the $\hat{z}$ basis for state readout.}
    \label{fig:experiment}
\end{figure}

We utilize an alternative technique, proposed in Ref.~\onlinecite{polloreno2022}, which employs spatial phase patterns applied to a large laser beam with an SLM. The technique can be used for local control of engineered spin-spin interactions, but here, as in Ref.~\onlinecite{polloreno2022}, we focus on the implementation of programmable rotations of the ion qubits. To turn the phase pattern into an ACSS pattern, a beam with flat wavefronts (``flat beam") is interfered with the beam with the imprinted phase pattern (``DM beam"). The interference between the two beams results in the Hamiltonian,
\begin{equation}
    \hat{H}=\sum_i\left\{U\sin\left[-\mu t+\varphi\left(\rho_i,\phi^l_i\right)+\psi\right]+\Delta_z\right\}\hat{\sigma}_i^z,\label{eq:H_DM_start}
\end{equation}
where $U$ is the maximum ACSS amplitude of the interfered beams, $\mu$ is the frequency difference between the beams, $\psi$ is a uniform relative phase between the beams, $\varphi\left(\rho_i,\phi_i^l\right)$ is the spatial phase deformation implemented by the SLM and described by the radial ($\rho_i$) and lab-frame azimuthal ($\phi_i^l$) coordinates for ion $i$, $\hat{\sigma}_i^z$ is the Pauli $z$ operator, and 
$\Delta_z=\Delta_z^{(\textrm{DM})}+\Delta_z^{(\textrm{flat})}$ is a uniform ACSS from the DM ($\Delta_z^{(\textrm{DM})}$) and flat ($\Delta_z^{(\textrm{flat})}$) beams individually. We write the phase pattern as $\varphi\left(\rho,\phi\right)=\varphi_\rho^m(\rho)\cos\left(m\phi^l\right)$. The azimuthally varying part of $\varphi$ can be written in terms of coordinates in the rotating frame of the ion crystal as $\cos\left[m\left(\phi^r-\omega_r t\right)\right]$. Here $\phi^r$ is the rotating frame azimuthal coordinate with rotation frequency $\omega_r$. Setting $\mu=m\omega_r$ and, assuming $t\geq2\pi/\omega_r$, with a rotating wave approximation that is exact for application durations equal to integer multiples of the rotation period, we obtain the Hamiltonian,
\begin{equation}
    \hat{H}\approx \sum_i\left\{UJ_1\left[\varphi_\rho^m\left(\rho_i\right)\right]\cos\left(m\phi_i^r-\psi\right)+\Delta_z\right\}\hat{\sigma}_i^z,\label{eq:H_DM}
\end{equation}
which is stationary in a frame rotating with the ion crystal \cite{polloreno2022}. If $m=0$ and $\mu=0$~kHz, the Hamiltonian instead is,
\begin{equation}
    \hat{H}\approx \sum_i\left\{U\sin\left[\varphi_\rho^0\left(\rho_i\right)+\psi\right]+\Delta_z\right\}\hat{\sigma}_i^z.\label{eq:H_DM_m=0}
\end{equation}
We note that Eqs.~\ref{eq:H_DM_start}-\ref{eq:H_DM_m=0} explicitly depend on the relative phase $\psi$ of the two beams; stabilizing this phase is therefore crucial.

More complicated patterns can be decomposed into a complete orthonormal basis set, such as Zernike polynomials, which are often used to describe optical aberrations and can be described as \cite{Born1981}, 
\begin{align}
    Z_n^m(\rho,\phi)&=R_n^m(\rho)\cos(m\phi) \nonumber\\
    Z_n^{-m}(\rho,\phi)&=R_n^m(\rho)\sin(m\phi). \label{eq:zern}
\end{align}
where $n$ and $m$ are the radial and azimuthal orders, respectively. Ref.~\onlinecite{polloreno2022} details how Zernike polynomials can be combined, either sequentially or in parallel, to achieve arbitrary patterns, including coherent individual addressing.

In this manuscript, we use this approach to imprint on crystals with up to 240 ions a variety of ACSS patterns that can be described with a single azimuthal order $m\leq$16. We compare projective measurements of the ion qubit populations with calculations of the populations using independent measurements of the applied ACSS pattern, finding population differences mostly $<$0.1, comparable to the calculated $\sim$0.1 population differences due to beam misalignments (see the Supplemental Material \cite{SM_arxiv}). Our work demonstrates that two co-propagating global beams can be used to achieve localized coherent addressing in a large, rapidly rotating crystal, with a pathway towards individual addressing when using an SLM with a larger number of actuators.

\textit{Experimental approach}--This work is performed in a Penning trap with a magnetic field of $\sim$4.46~T, oriented along $\vec{z}$, which we refer to as the axial direction [see Fig.~\hyperref[fig:experiment]{\ref*{fig:experiment}(a)}]. We use an ion crystal rotation frequency of $\omega_r/(2\pi) = 180$~kHz, which is controlled by a rotating wall potential \cite{huang1998} obtained by applying oscillating potentials to the segmented electrodes at the vertical trap center. This frequency yields a single-plane crystal with a diameter $<250~\mu$m for the ion numbers used for this study ($\lesssim250$ ions). The details of the atomic structure and transitions driven are shown in the End Matter. Our qubit is composed of two ground state Zeeman levels, which at 4.46~T are separated by $\sim$124~GHz. We Doppler cool with two beams, one parallel and one perpendicular to the magnetic field, and perform state readout using fluorescence from the same atomic transition. Ion fluorescence is collected parallel to the magnetic field, and sent to a camera (not shown), which records photon arrival times in addition to the standard position coordinates \cite{Roentdek}. Combined with rotating wall control of $\omega_r$, this enables imaging in the rotating frame of the crystal and therefore determination of the state of individual ions. Similar time stamping techniques for rotating frame imaging were used in Refs.~\onlinecite{britton2012,bohnet2016,wolf2024,mcmahon2024,jee2026,mitchell2001}.

In this initial demonstration, we use a deformable mirror (DM) with 137 actuators arranged in a square grid with a pitch $\Lambda = 300~\mu\textrm{m}$~\cite{BMC_DM}. The DM surface consists of an aluminum-coated membrane, which we have measured to have $>$80$\%$ reflectivity at 313~nm \cite{SM_arxiv}. Electrostatic actuators locally deform the surface by distance $d_j$ for actuator $j$, resulting in a phase shift of $\varphi_j=4\pi d_j/(313~\mathrm{nm})$.
\begin{figure*}[t]
    \centering
    \includegraphics[width=\textwidth]{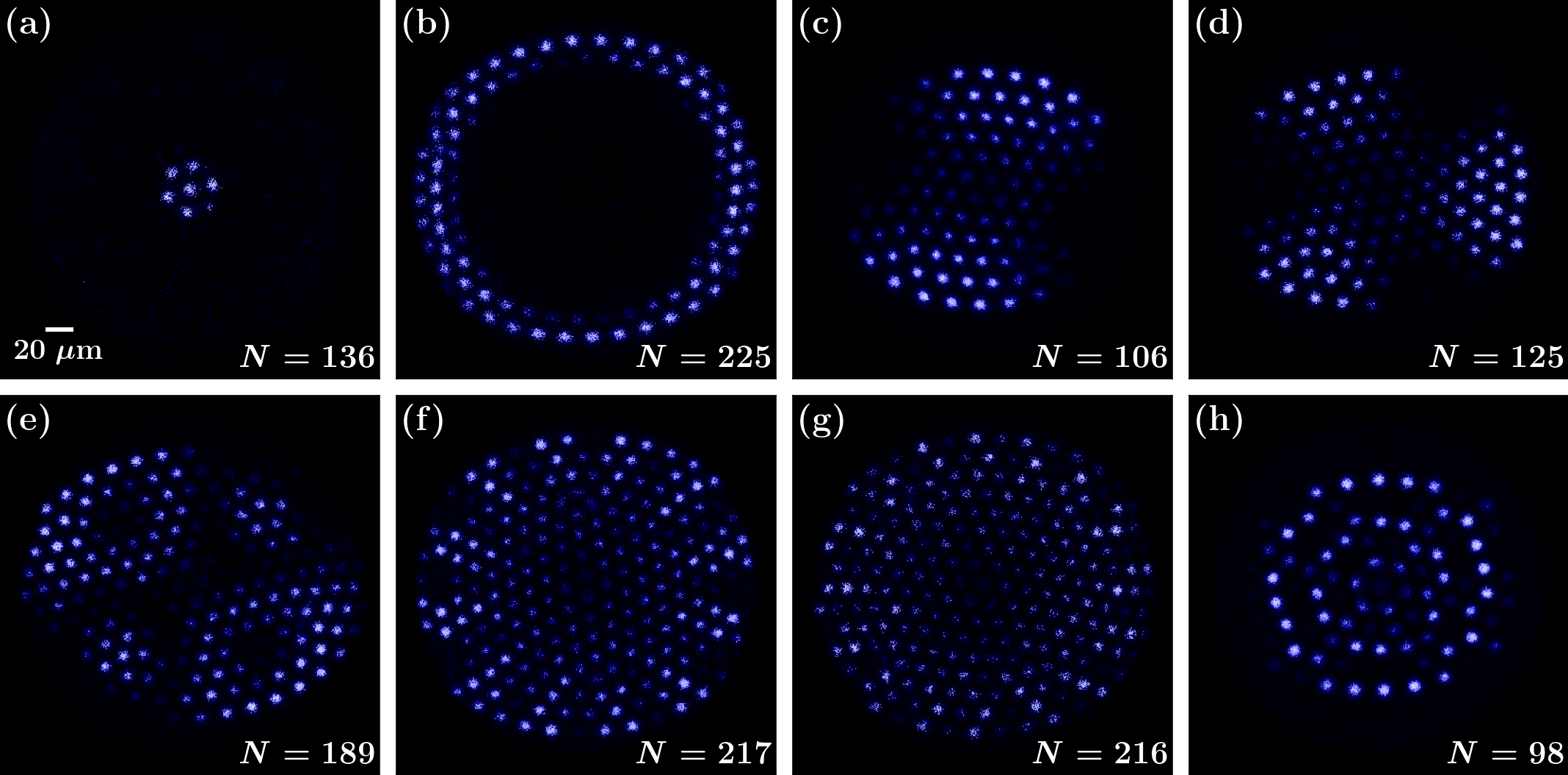}
    \caption{Example ion fluorescence images from the first detection step. Brighter ions correspond to a larger bright fraction. (a)-(b) Rings--$R=0$ and $R=4$. For both rings, $\varphi_{max}=\pi,\;t=t_{\pi/4}$. (c) $Z_n^m=Z_2^2,\;\varphi_{max}=\pi,\;t=2t_{\pi/4}$. (d) $Z_n^m=Z_5^3,\;\varphi_{max}=2\pi,\: t=2t_{\pi/4}$. (e) $Z_n^m=Z_4^2,\;\varphi_{max}=2\pi,\;t=4t_{\pi/4}$. (f)-(g) Azimuthally varying rings--$R=4,\;m=16$ and $R=4,\;m=12$. For both azimuthally varying rings, $\varphi_{max}=0.58\pi$ and $t=2/.58\times t_{\pi/4}$. (h) Linear radial gradient, $2\varphi_{max}/\rho_{max}=2\pi,\;\rho_{max}=6,\;t=t_{\pi/4}$.}
    \label{fig:patterns}
\end{figure*}

The DM and flat beams are tuned approximately 3~GHz from the $\ket{S_{1/2},m_J=-1/2}\leftrightarrow\ket{P_{3/2},m_J=-3/2}$ transition (see the End Matter). This relatively small detuning was chosen to balance minimizing beam powers and off-resonant scattering. However, this detuning means that $\Delta_z\neq0$ in Eqs.~\ref{eq:H_DM_start}-\ref{eq:H_DM_m=0}. This term is cancelled with a spin-echo sequence.

The parallel Doppler cooling, flat, and DM beams are sent through a telescope, which images the DM surface onto the ions with a lateral demagnification of about 10, resulting in a beam diameter of $\sim350~\mu$m. The beams propagate parallel to $\vec{B}$ as shown in Fig.~\hyperref[fig:experiment]{\ref*{fig:experiment}(a)}. A beam sampler is used to pick off $\sim$10$\%$ of the beams. The picked-off light is sent to a CMOS camera placed on the optical table with a replicated setup of the optics used to send the beams to the ions. More details on these optics are provided in the End Matter and shown in Fig.~\ref{fig:dm_optics}. The picked-off beam is critical for fine-tuning the relative flatness of the wavefronts of the DM and flat beams and for independently measuring the ACSS pattern. Periodically, we apply small corrections to a nominally flat DM surface to reduce the average deviation of the relative wavefronts of the two beams to $\lesssim2$~nm \cite{SM_arxiv}. All pattern displacements are added to this adjusted surface.
\begin{figure*}[t]
    \centering
    \includegraphics[width=\textwidth]{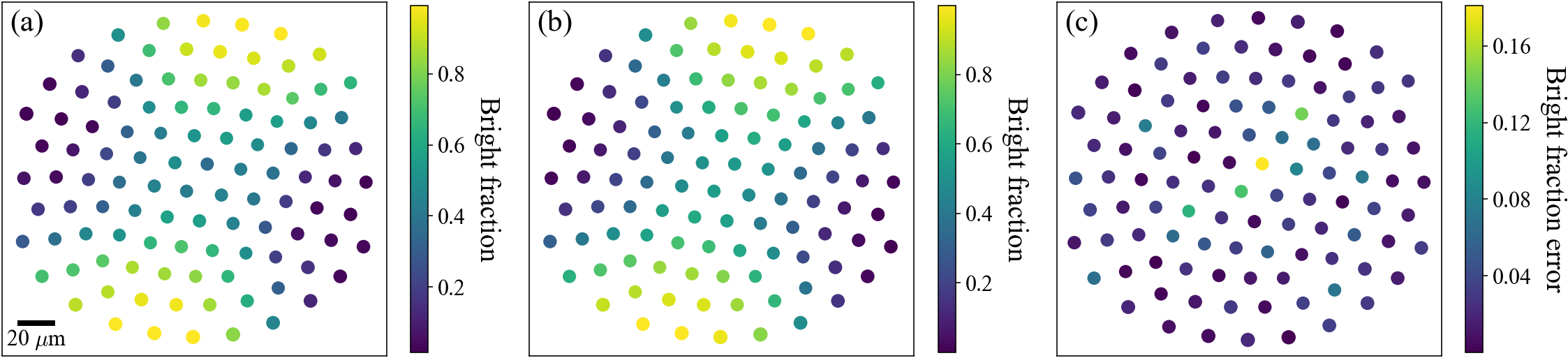}
    \caption{Plots of bright fractions and bright fraction errors at ion locations for the $Z_2^2$ pattern shown in Fig.~\hyperref[fig:patterns]{\ref*{fig:patterns}(c)}. (a) Measured ion bright fractions. (b) Predicted ion bright fractions. (c) Bright fraction error. The average error $\epsilon_{avg}$ is $\sim$0.04.}
    \label{fig:bright_fraction_plots}
\end{figure*}

We apply the ACSS pattern in a spin-echo sequence as shown in Fig.~\hyperref[fig:experiment]{\ref*{fig:experiment}(b)}, where global microwave rotations are driven with a 124~GHz source. The DM and flat beams are switched on in the two arms of the spin echo for a time $t\sim10-120~\mu\textrm{s}$ with relative phase $\psi$ and beatnote $\mu$. This pulse sequence allows for the accumulation of spin precession due to the spatially-varying ACSS $\propto U\varphi\left(\rho_i,\phi_i^r\right)$ but cancels $\Delta_z$. A final microwave $\pi/2$ pulse projects the spin precession onto the Bloch sphere $z$ axis, which we can directly measure. This pulse sequence maps ACSS patterns onto ion qubit spin states patterns. After every five repetitions of this sequence, we switch the DM surface to the flat configuration and measure the relative phase between the two beams using the ions \cite{SM_arxiv}. We then adjust the phase of the RF drive of the acousto-optic modulator that is also used to switch and adjust the frequency of the flat beam. This phase tracking process mitigates any slow drifts in the relative phase $\psi$ that would otherwise inevitably occur, such that the phase stability does not limit the length of time over which we can imprint a pattern.

Our projective spin state detection protocol consists of recording a rotating frame image for 25~ms, a microwave $\pi$ pulse to flip ions that were dark to bright and vice versa, followed by another image with an exposure time of 25~ms. We subtract the counts for each ion in the second image from the first; a positive number indicates a bright ion \cite{SM_arxiv}. 

\textit{Results}--We apply four categories of patterns with the DM actuator phases $\varphi$ set with a variety of azimuthal orders: (i) rings where $m=0$ and
$\varphi=\varphi_{max}\delta_{\rho'R}$, (ii) azimuthally varying rings where $\varphi=\varphi_{max}\delta_{\rho'R}\cos\left(m\Phi_{DM}\right)$, (iii) Zernike polynomials where $\varphi=\varphi_{max}Z_n^m\left(\rho_{DM},\Phi_{DM}\right)$, and (iv) linear radial gradients where $\varphi=\varphi_{max}\left(1-2\rho_{DM}/\rho_{max}\right)$. In these expressions, $\varphi_{max}=\max\left(\left\vert\varphi_j\right\vert\right)$, $R$, an integer, is the radius of the ring in units of the DM pitch $\Lambda$, $\rho_{max}$ is an integer that is the maximum radius in units of $\Lambda$ of the DM actuators that are set for the pattern, $\rho_{DM}$ and $\Phi_{DM}$ are the DM actuator radial and azimuthal coordinates, respectively, $\lfloor\cdot\rceil$ indicates rounding to the nearest integer, $\rho'=\lfloor\rho_{DM}/\Lambda\rceil$, and $\delta_{R_1 R_2}$ is the Kronecker delta. $\rho_{max}$ is typically set larger than the crystal radius accounting for a demagnification of 10.

Qualitative ion images obtained after the application of different ACSS patterns are shown in Fig.~\ref{fig:patterns}. These images are generated by adding the counts from the rotating frame images obtained in the first 25~ms detection period over multiple repetitions of the sequence shown in Fig.~\hyperref[fig:experiment]{\ref*{fig:experiment}(b)}. The arm times $t$ are determined as multiples of the time to drive a $\pi/4$ rotation, $t_{\pi/4}$, with flat wavefronts \cite{SM_arxiv} and $\psi$ set to maximize the ACSS amplitude. 

For $m=0$ ring patterns, $\psi$ is set such that the ions in that ring are bright. The brightness of the ions shown in Fig.~\ref{fig:patterns} is an indicator of the bright fraction but not a quantitative measurement due to significant variations in quantum efficiency across the camera. In Figs.~\hyperref[fig:patterns]{\ref*{fig:patterns}(a)-(b)}, we show rings with $R=0$ (raised center actuator) and $R=4$, respectively, and $\varphi_{max}=\pi$. The ring with $R=0$ illustrates our current addressing resolution. With inter-ion spacings of $\sim$15~$\mu$m, we observe that all ions except the center ion and a single ring of ions around it are dark, indicating an approximate 30~$\mu$m resolution. This is consistent with the $\sim$$\times$10 demagnification and 300~$\mu$m DM pitch.

In Figs.~\hyperref[fig:patterns]{\ref*{fig:patterns}(c)-(g)}, we show patterns with $m\neq0$. For these patterns we set $\mu=m\omega_r$, resulting in a pattern that is stationary in the rotating frame. With Zernike polynomials, we show patterns with $m\leq4$. This limit is largely set by the resolution of the deformable mirror and the fact that higher order polynomials typically have more features around the edge of the pattern, which is outside the boundary of the ion crystal. Finally, we demonstrate patterns with $m>4$ with azimuthally asymmetric rings, since rings further from the center can implement higher order patterns due to the larger number of actuators. 
For the azimuthally asymmetric rings, $\varphi_{max}$ was set to maximize $J_1\left(\varphi_{max}\right)$ in the Hamiltonian in Eq.~\ref{eq:H_DM}. The times $t$ were then chosen such that $\pm2UJ_1\left(\varphi_{max}\right)t=\pm\pi/2$. Finally, in Fig.~\hyperref[fig:patterns]{\ref*{fig:patterns}(h)}, we show a linear radial gradient pattern. These images demonstrate the large number of ions we can use for these experiments, the localization of the ions, and that our resolution remains similar for ions at the edges of the crystal. 

To evaluate the performance of the patterned addressing quantitatively, we compare the bright state fraction of each ion obtained from the projective spin state measurements with a calculation based on the phase and amplitude maps generated with the images recorded on the CMOS camera. An example of this comparison for the pattern shown in Fig.~\hyperref[fig:patterns]{\ref*{fig:patterns}(c)} is shown in Fig.~\ref{fig:bright_fraction_plots}. We compute the difference $\epsilon_i$, which we refer to as the error [Fig.~\hyperref[fig:bright_fraction_plots]{\ref*{fig:bright_fraction_plots}(c)}], between the measured [Fig.~\hyperref[fig:bright_fraction_plots]{\ref*{fig:bright_fraction_plots}(a)}] and the calculated [Fig.~\hyperref[fig:bright_fraction_plots]{\ref*{fig:bright_fraction_plots}(b)}] bright fractions for each ion $i$ \cite{SM_arxiv}. The average error for a pattern is calculated as $\epsilon_{avg}=(1/N)\sqrt{\sum_i\epsilon_i^2}$. We find values of $\epsilon_{avg}$ mostly $<0.1$. A global $\pi/2$ pulse driven with the flat and DM beams with flat wavefronts results in errors of $\sim0.01$ \cite{bullock_thesis}. 

\textit{Discussion and conclusion}--The alignment of the replicated optical setup for the CMOS camera will not exactly match that in the bore of the magnet, resulting in expected errors of order 0.1 for some patterns~\cite{SM_arxiv}. Additionally, projection noise can contribute significantly (0.05 for 100 repeated measurements, which is close to the average number of measurements analyzed for a scan). An analysis in the Supplementary Material~\cite{SM_arxiv} shows a faster increase in the error rate with the arm time used for the ACSS application than expected from spontaneous emission. Errors from misalignments are expected to be worse at longer arm times. Overall, the finding that, despite these potentially significant error sources, $\epsilon_{avg}<$0.1 for most patterns validates the utility of employing a SLM for localized control of rotating ion crystals in a Penning trap.

More complex patterns can be achieved by applying patterns with multiple radial and azimuthal orders. Individual ion addressing will require a maximum radial order $n_{max}$ of $\sim$40, resulting in a minimimum actuator number of order $n_{max}^2=1600$ \cite{polloreno2022}. UV-compatible SLMs with sufficient actuator numbers have recently been demonstrated~\cite{ammenwerth2025}. Further demagnification will also likely be required, motivating the use of an objective lens for imaging the SLM surface onto the ion crystal. We note that high motional mode frequencies, which result in high rotation frequencies, are advantageous for quantum information processing. The technique discussed here is suitable for generating rapidly rotating ACSS patterns since the rotation of the pattern is set only by the frequency difference between the DM and flat beams. In comparison, the timing requirements for addressing using tightly-focused beams become more stringent as the rotation frequency increases.

In addition to the coherent addressing demonstrated here, employing SLMs for local control of engineered spin-spin and spin-motion interactions provides a powerful tool for enhancing quantum simulation possibilities with trapped ion crystals. Combined with an axial spin-dependent force, spatially inhomogeneous Hamiltonians that couple the spin degrees of freedom with the axial COM motional mode of the ions can be engineered, as is required, for example, for the quantum simulation of $p+ip$ superconductivity proposed in Ref.~\onlinecite{shankar2022}. Finally, because the technique demonstrated here enables the application of ACSS gradients in the plane of the ion crystal, it can also be used to couple the ion qubit degrees of freedom to the in-plane motional modes (see the End Matter for a discussion of these modes). This coupling can be used, for example, for thermometry of the in-plane motion through measurements of spin dephasing \cite{sawyer2014} or, with the application of sufficiently strong gradients, to implement quantum simulation employing the in-plane modes, which have interesting properties such as chiral asymmetry \cite{hawaldar2024}. 

\textit{Acknowledgements}--We thank A. M. Rey and D. Leibfried for useful discussions and comments on this manuscript, and X.-P. Huang for discussions on imaging the DM surface to the ion crystal. 
This work was supported by AFOSR FA9550-25-1-0080, the
DARPA ONISQ program, NIST, the U.S. Department of Energy, Office of Science, National Quantum Information Science Research Centers, Quantum Systems Accelerator. JFL acknowledges support from an Intelligence Community Postdoctoral Research Fellowship.
\\
\\
This document has not been peer reviewed but has been cleared by NIST for release.

\section*{End matter}
\begin{figure}
    \centering
    \includegraphics[width=\linewidth]{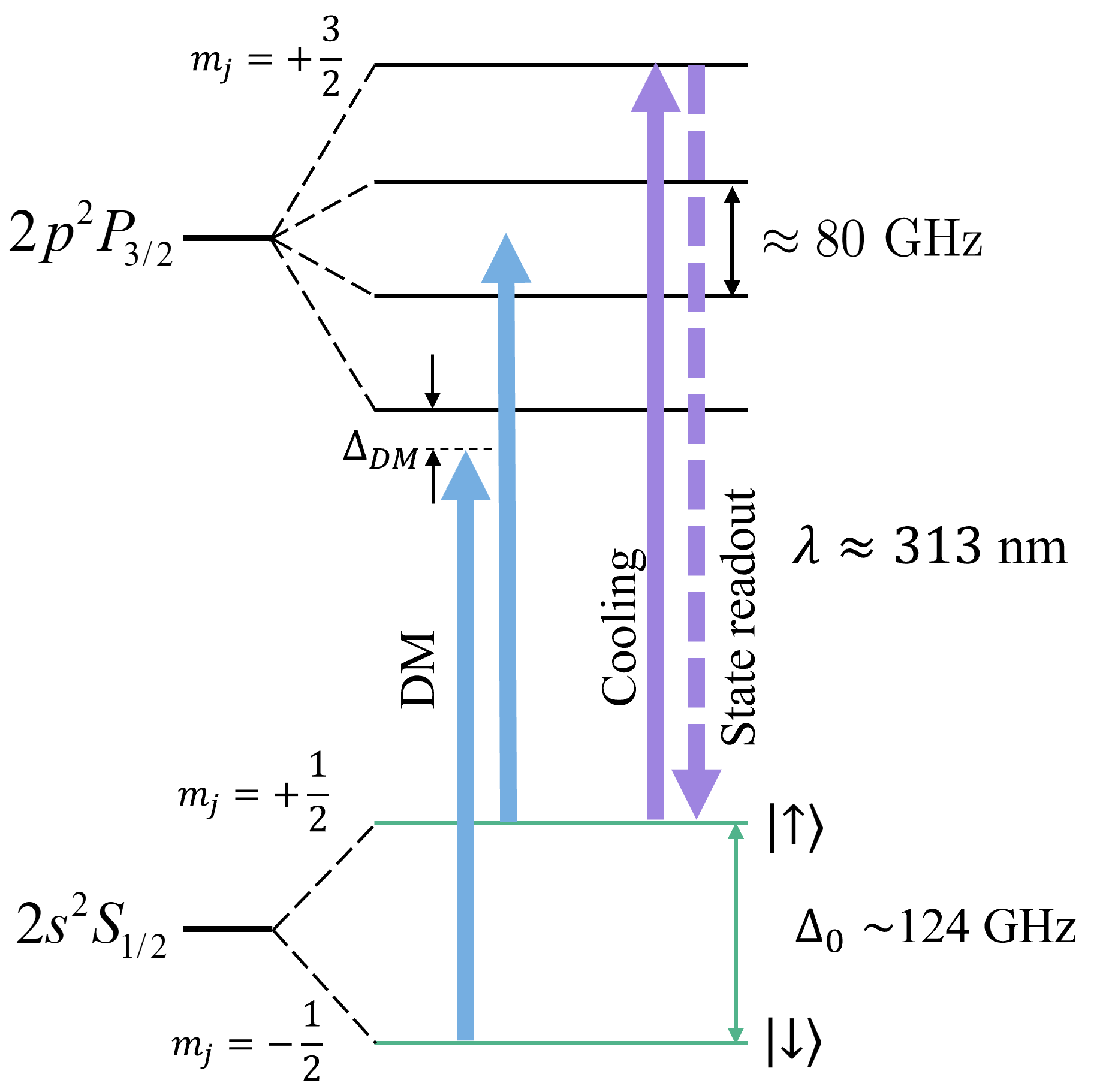}
    \caption{$^9$Be$^+$ atomic structure at 4.5~T with relevant laser beams. $\Delta_{DM}$ indicates the $\sim$3~GHz detuning from the $\vert\downarrow\rangle\leftrightarrow\left\vert P_{3/2},m_J=-3/2\right\rangle$ transition.}
    \label{fig:energy_levels}
\end{figure}
\textit{Experimental setup}--At high magnetic fields, the $^9$Be$^+$ ions are pumped to the $m_I=+3/2$ nuclear spin state during Doppler cooling, resulting in an effective atomic structure of a spin-zero nucleus [see Fig.~\ref{fig:energy_levels}] \cite{itano1981a}. The qubit is defined as $\ket{\downarrow}$$=$$\ket{S_{1/2},m_J=-1/2}$ and $\ket{\uparrow}$$=$$\ket{S_{1/2},m_J=+1/2}$. The DM and flat beams, shown in blue, are detuned $\sim$3~GHz from the $\ket{\downarrow}\leftrightarrow\ket{P_{3/2},m_J=+3/2}$ transition. Doppler cooling and state readout are performed using the $\ket{\uparrow}\leftrightarrow\ket{P_{3/2},m_J=+3/2}$ cycling transition. Microwaves at 124~GHz drive global qubit rotations. The large Zeeman splittings of 124~GHz in the $S_{1/2}$ manifold and 80~GHz in the $P_{3/2}$ manifold enable long detection times with minimal undesired repumping.

\begin{figure*}[t]
    \centering
    \includegraphics[width=.9\textwidth]{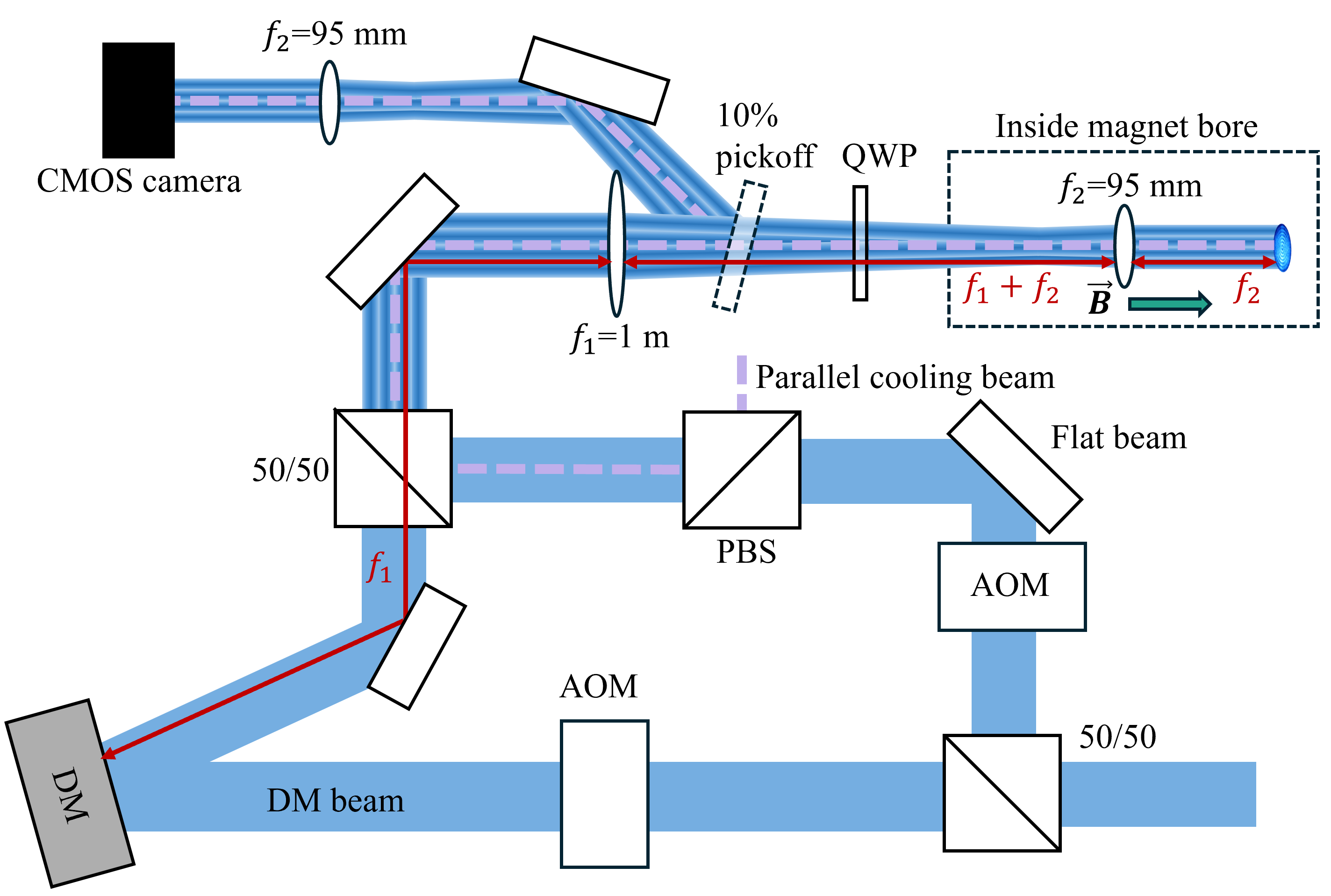}
    \caption{Beam paths for the parallel cooling (dashed purple), DM (blue), and flat (blue) laser beams (not to scale). The DM and flat beams are shown in solid blue before they are combined and in a blue gradient after their combination, indicating the imprinted phase gradient. A pickoff directs 10$\%$ of the light to a CMOS camera on the optical table, while the remaining light propagates to the ions parallel to $\vec{B}$, which is indicated with the green arrow. Two lenses with focal lengths $f_1=1$~m and $f_2=95$~mm image the DM surface onto the CMOS camera and the ion crystal. Red arrows indicate the distances between optical elements--the DM to the first lens ($f_1$), the first lens to the second ($f_1+f_2$), and the second lens to the ions ($f_2$). In the beam path to the ions, the $f=95$~mm lens is located inside the bore of the superconducting magnet. Unlabeled rectangles represent mirrors.}
    \label{fig:dm_optics}
\end{figure*}

A schematic of the Penning trap used in this work is shown in Fig.~\hyperref[fig:experiment]{\ref*{fig:experiment}(a)}. Static voltages applied to the cylindrical electrodes provide a confining potential in the direction of the magnetic field ($z$ axis). The radial confinement arises from the Lorentz force generated by the ion crystal rotation through the magnetic field. The mutual Coulomb repulsion of the ions and the confining potential together result in $3N$ motional modes for $N$ ions. For a planar crystal, the modes are divided into three branches--the low frequency $\exb$ modes, the high frequency cyclotron modes, and the drumhead axial modes, which have intermediate frequencies \cite{wang2013}. The axial and cyclotron modes are cooled with Doppler cooling beams sent parallel and perpendicular to the magnetic field respectively. The perpendicular cooling beam is offset from the center of the crystal, resulting in an intensity gradient which cools the $\exb$ modes \cite{torrisi2016,johnson2024}. The cooling of the $\exb$ modes is very inefficient with this technique, so we add in a mode coupling drive to couple the $\exb$ modes with the well-cooled cyclotron modes. The details of this scheme will be discussed in a publication that is currently in preparation. 

\textit{DM optics details}--Figure~\ref{fig:dm_optics} provides a more detailed schematic of the optical setup for the DM, flat, and parallel Doppler cooling beams. A single laser beam is split into two beams with approximately equal amplitudes. Each beam separately passes through an AOM, which is used to shift the frequency of the beam and switch the beam on or off. Additionally, the AOM in one of the beams is used to adjust the beatnote frequency $\mu$ and the relative phase between the beams $\psi$. Lenses in both beam paths (not shown) are used to set the waist of both beams to approximately the size of the 3.6~mm diameter of the DM active area. The angle of incidence of the DM beam on the DM is about $7^\circ$, which is the smallest angle that is feasible with the available space on the optical table. The parallel Doppler cooling beam is combined with the flat beam on a polarizing beam splitter (PBS). Both beams are then overlapped with the DM beam on a 50/50 beam splitter. A quarter wave plate (QWP) rotates the polarization to circular polarization resulting in $\hat{\sigma}^-$ polarization for the DM and flat beams and $\hat{\sigma}^+$ for the cooling beam. This configuration simultaneously maximizes the ACSS from the DM and flat beams and optimizes the cooling beam polarization for the $\ket{\uparrow}\leftrightarrow\ket{P_{3/2},m_J=+3/2}$ transition, thereby maximizing the signal-to-noise ratio during state readout by increasing the ion scattering rate while keeping the background scattering rate fixed. All three beams co-propagate to the ions parallel to $\vec{B}$. The co-propagation of the DM and flat beams results in improved phase stability since they will mostly be affected only by common-mode path length fluctuations. Residual phase fluctuations can arise from modulation of the beam paths when separated, so we aim to minimize the area enclosed between the two beams.

A telescope where the first lens has focal length $f_1=1$~m and the second has focal length $f_2$=95~mm images the DM surface onto the ions when the distance from the DM to the first mirror is approximately $f_1$, the distance between the two lenses is approximately $f_1+f_2$, and the final lens is a distance of $f_2$ from the ions. This short distance requires placing the lens inside the room-temperature bore of our superconducting magnet. This final distance is the easiest to determine accurately, since when the 1~m focal length lens is removed, we are able to observe a clear focus of the parallel cooling beam on the ions and minimize the resulting spot size via a mechanical adjustment setup in the bore of the magnet. The two lenses image the DM surface at the plane of the ions and demagnify the beams and the spatial phase modulation pattern by a factor of about 10. This results in an overall beam diameter for both the flat and DM beams of about 350~$\mu$m at the ions, and the size of the area spanned by each DM actuator is about $30~\mu\textrm{m}\times30~\mu\textrm{m}$.

A 10$\%$ pickoff is placed between the two lenses in the telescope. The picked-off beam is directed to a replica setup of the second telescope lens (focal length $f_2$). A CMOS camera is placed a distance $f_2$ away from the lens. The images obtained with the camera are used for compensating beam aberrations and for comparing the imprinted and expected phase patterns. The details of this comparison are provided in the Supplemental Material \cite{SM_arxiv}.

\bibliography{References}

\begin{thebibliography}{45}
\bibitem [{pro()}]{product}%
  \BibitemOpen
  \href@noop {} {}\bibinfo {note} {Certain commercial equipment, instruments, or materials are identified in this paper in order to specify the experimental procedure adequately. Such identification is not intended to imply recommendation or endorsement by NIST, nor is it intended to imply that the materials or equipment identified are necessarily the best available for the purpose.}\BibitemShut {Stop}%
\end{thebibliography}
\clearpage

\setcounter{section}{0}
\setcounter{figure}{0}
\setcounter{equation}{0}
\renewcommand{\thefigure}{S\arabic{figure}}
\renewcommand{\theequation}{S\arabic{equation}}

\begin{widetext}
\begin{center}
\Large\textbf{Supplemental material for\\Localized control of large ion crystals in a Penning trap \\using a spatial light modulator}
\end{center}
\end{widetext}
\setcounter{section}{0}
\renewcommand{\thesection}{S\arabic{section}}
\section{Deformable mirror characteristics}
Here, we provide some key specifications for the deformable mirror~\cite{BMC_DM}.
\begin{itemize}
    \item 83$\%$ reflectivity at 313~nm
    \item Actuator number: 137
    \item Actuator size: 300~$\mu$m
    \item Maximum actuator displacement: 1.5~$\mu$m
    \item Settling time: $<50~\mu$s
\end{itemize}

\begin{figure*}
    \centering
    \includegraphics[width=.8\textwidth]{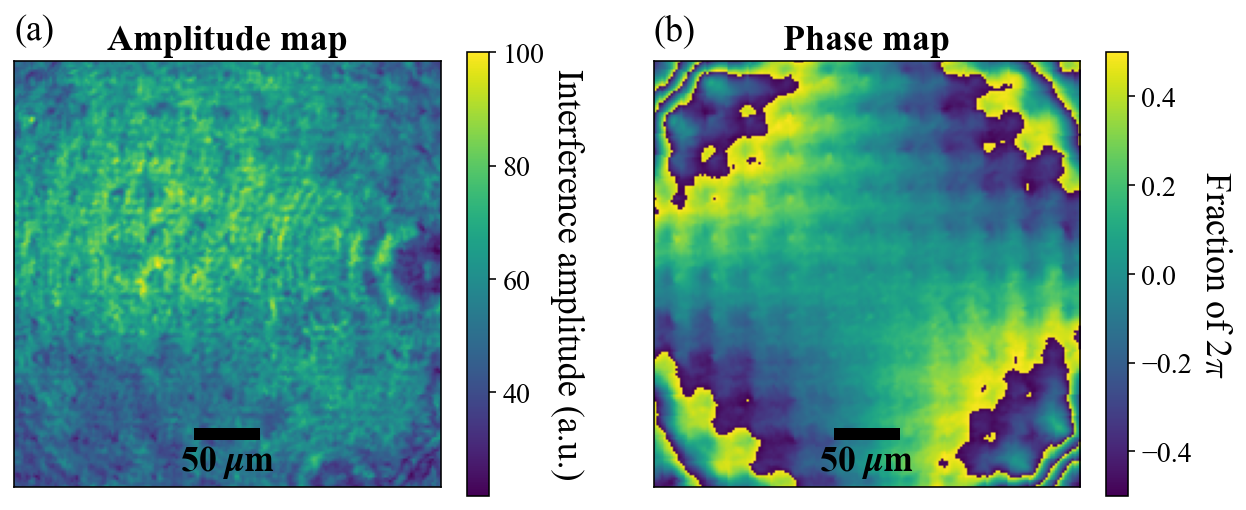}
    \caption{Example amplitude and phase maps for a $Z_2^2,\,\varphi_{max}=2\pi$ pattern. (a) Amplitude map obtained by fitting CMOS images. The average amplitude at a radius of 150~$\mu$m, which is larger than the radius of the largest crystals used in this work, is lower than the amplitude at the center by about 20$\%$. (b) Phase map obtained by fitting CMOS images. The phase map is not unwrapped, which allows easier visualization of the pattern.}
    \label{fig:pm_am}
\end{figure*}

\section{Calibrations and preparation}\label{sec:calibrations}

\subsection{CMOS camera images for wavefront compensation and measuring expected patterns}\label{sec:CMOS}
We record a series of images on the CMOS camera with the DM and flat beams set to identical frequencies while varying the relative phase between the beams $\psi$. For each pattern, we fit each pixel to a sinusoidal function, yielding an amplitude, offset, and phase. From these fits, we generate an amplitude map which accounts for the local beam intensity, and a phase map, which depends on the path length differences induced by the DM actuators. A full $\varphi_j=2\pi$ phase shift corresponds to an actuator displacement of actuator $j$ of $d_j=\lambda/2=156$~nm. Example amplitude and phase maps for $Z_2^2,\,d=156$~nm are shown in Fig.~\hyperref[fig:pm_am]{\ref*{fig:pm_am}(a)} and Fig.~\hyperref[fig:pm_am]{\ref*{fig:pm_am}(b)}, respectively. This technique is used both for calculating the expected imprinted patterns and for initial optimization of the DM surface. 

\begin{figure*}[t]
    \centering
    \includegraphics[width=\textwidth]{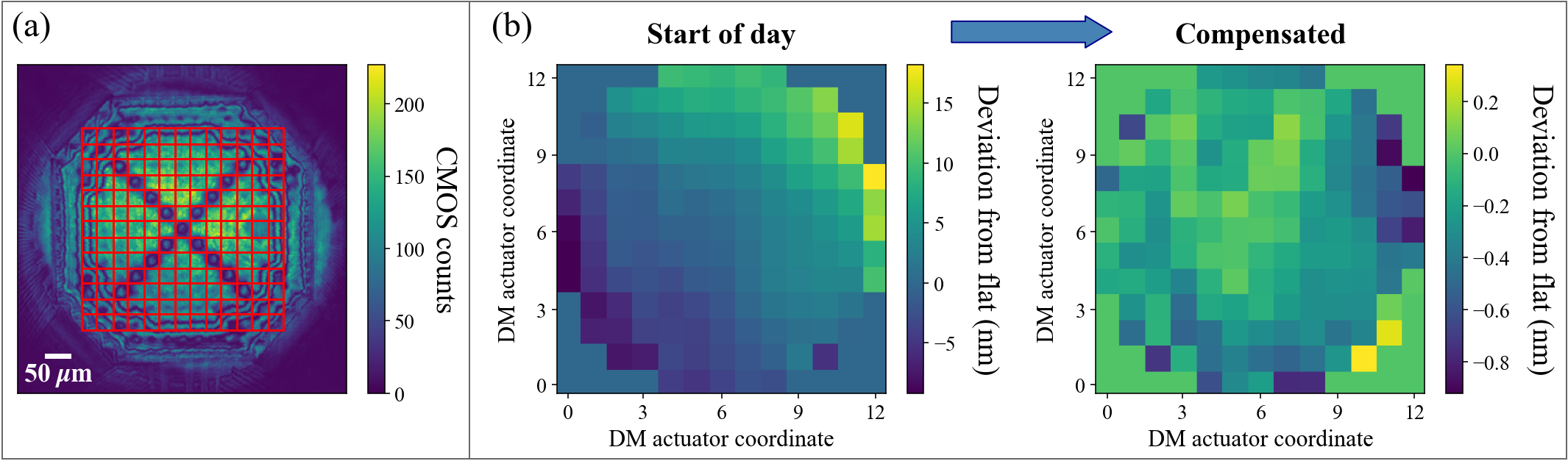}
    \caption{Centering and wavefront flattening process. (a) An `X'-shaped actuator pattern is set. A grid, which determines the center of the DM pattern for analysis, is superimposed, showing the alignment of the center grid square with the center actuator. The size of the squares is set to match as closely as possible the size of the DM actuators displaced in order to apply the `X' pattern, especially near the center. (b) Displacements averaged over the area of each actuator measured with CMOS images before and after compensation. Each pixel corresponds to a square of the grid in (a), and therefore one DM actuator. An average phase is calculated for each pixel. Due to drifts in beam alignments, a tilt is observed in the first plot, with a maximum offset of $\sim$20~nm (or a phase shift of $\sim$0.26$\pi$). The DM actuators are adjusted to decrease these deviations from flat wavefronts. After several adjustments, the maximum average deviation from a flat wavefront of any actuator is $\lesssim$1~nm.}
    \label{fig:flattening}
\end{figure*}

Ideally, both beams would have perfectly flat wavefronts before the DM and would be perfectly co-propagating after being combined. In this case, the phase map extracted from the CMOS images should be uniform when the DM actuators are set to $\varphi_{j}=0$ for all $j$. Even a small amount of misalignment through optics or in the relative beam paths, however, can introduce deviations from the ideal case. By iteratively observing the deviations from flat wavefronts on the CMOS camera and adjusting the DM actuators accordingly, we can compensate for these small imperfections. Before beginning this compensation process, we apply a pattern where individual actuators are raised to form an `X' shape and record an image of the resulting interference pattern on the CMOS camera. In software, we align a grid pattern with the `X' to determine the center of the DM surface and the size and locations of the actuators [see Fig.~\hyperref[fig:flattening]{\ref*{fig:flattening}(a)}]. We then set the DM surface to nominally flat $\left(\varphi_j=0\right)$ for all $j$, which would ideally generate a completely constant interference pattern. We record a set of images with varying $\psi$ as discussed above and generate a phase and amplitude map. For the phase map, we average over each actuator area as determined by the previously aligned grid to calculate a compensatory displacement for each actuator. As long as the initial deviations from the ideal phase are $\ll\pm\pi$, which we can achieve relatively straightforwardly with manual beam alignment, 5-10 iterations of observing the deviations from flat wavefronts and adjusting the actuators typically result in our desired compensated flat surface. A representative example of the averaged deviations from flat wavefronts at the beginning of a day is shown in the left plot in Fig.~\hyperref[fig:flattening]{\ref*{fig:flattening}(b)}, where the maximum deviations from flat wavefronts are $<20$~nm, a small fraction of a $\pi$ phase shift. We adjust each actuator to decrease the deviation and iterate until we reach a maximum deviation $\lesssim$2~nm, often $\lesssim$1~nm [right plot in Fig.~\hyperref[fig:flattening]{\ref*{fig:flattening}(b)}]. On the timescale of a day, we typically see drifts in the flatness of the phase pattern corresponding to a change in the relative tilt of the beams of $\ll$1 mrad. This tilt can be easily compensated with the DM. Manual adjustments are typically needed only every couple of weeks.

\subsection{Ion calibrations}\label{sec:ion_calibrations}

To determine the arm time $t$ for applying patterns, we set the DM to its compensated flat surface and observe spin precession induced by the DM and flat beams in a Ramsey-type sequence with a spin echo. We begin by rotating all spins to the equator of the Bloch sphere with a microwave $\pi/2$ pulse. We then apply the DM and flat beams for two arms with arm times $t$ with a microwave $\pi$ pulse in between. A microwave $\pi/2$ pulse after the second arm projects the resulting spin precession onto the $z$ axis of the Bloch sphere. We read out global fluorescence and observe clear flopping between the bright and dark states. We periodically perform the same phase tracking procedure as is used for imprinting patterns to set $\psi$ to maximize the constructive interference, which in turn maximizes the AC Stark shift (ACSS) amplitude $U$. The spin precession data are fit to a sinusoidal function, which determines the time required to drive a $\pi/2$ $\hat{\sigma}^z$ rotation. Typically, this is in the range of 20-25~$\mu$s (4-5 crystal rotation periods). This time is divided by two to obtain the $\pi/4$ time, $t_{\pi/4}$. With two arms, a time $t=t_{\pi/4}$ for each arm would result in a $\pi/2$ $\hat{\sigma}^z$ rotation with a flat pattern of all ions. All arm times $t$ are set as multiples of $t_{\pi/4}$.

For each pattern we imprint, we desire deterministic control of $\psi$. Despite the beams' co-propagation, we observe slow drifts in the relative phase, presumably from the small area where the paths are separated. Thus, we periodically measure the current relative phase of the beams. To do this, we set the DM surface to the compensated flat surface and scan $\psi$ by adjusting the phase of the RF drive for the acousto-optic modulator (AOM) in the path of the flat beam. We apply the beams for the time it takes to drive a $\pi/2$ rotation when $\psi$ is set to maximize the ACSS. As $\psi$ is scanned, we observe a sinusoidal oscillation in the bright fraction corresponding to the amount of spin precession and therefore the effective ACSS for that value of $\psi$. This allows us to apply an offset phase such that we can obtain the desired relative phase at the beginning of each scan when we're imprinting patterns. 

As mentioned in the main text, we also track and stabilize this relative phase during the imprinting of the ACSS patterns. Instead of doing a full scan of $\psi$ as described in the previous paragraph, we use three points at the sides and minimum of the phase curve to adjust the phase offset and compensate for any slow drifts by adjusting the phase of the RF AOM drive for the flat beam. The efficacy of the phase stabilization is best demonstrated by the highest $m$ patterns, where deviations in the set phase greater than $10^\circ$ would rotate the pattern and thus smear out the features.

The centration of the pattern on the ion crystal is important for high-fidelity pattern imprinting. More details of our centering procedure will be discussed in a future publication, but we include a basic summary here. To optimize the alignment of the center of the pattern to the center of the ion crystal, we begin by applying a linear radial gradient pattern, where the displacement of an actuator on the DM linearly increases with the actuator distance from the DM center. This pattern can be described as 
$\varphi\left(\rho_{DM}\right)=\varphi_{max}\left(1-2\rho_{DM}/\rho_{max}\right)$, where $\rho_{DM}$ is the radial coordinate of the DM actuator, $\varphi_{max}$ is the maximum phase shift, and $\rho_{max}$ is an integer value for the maximum radius of actuators displaced in units of the DM pitch $\Lambda$, which is typically larger than the ion crystal radius. The radial gradient enables a relatively simple relationship between the spatial extent of the ions due to thermal fluctuations of the in-plane motion and spin dephasing that arises as the ions move through varying ACSSs \cite{bullock_thesis}. We set $\mu$ close to the frequency of a motional mode. In this case, we typically use the cyclotron center-of-mass mode with frequency $\omega_c$. A centering misalignment results in sidebands with a modulation index determined by the rotation frequency and the size of the misalignment relative to the wavelength of the radial gradient. The power in the main tone is completely removed at a zero of a Bessel function corresponding to a 7~$\mu$m misalignment. On the other hand, if we set $\mu=\omega_c\pm\omega_r$, there will be a nonzero sensitivity to the motion if the pattern is off-center. Therefore, we can clearly see a dependence on the sensitivity to spin dephasing with an alignment error on the scale of 1-2~$\mu$m. To optimize the alignment, we translate the 1~m lens [see Fig. 5 in the End Matter of the main text] to minimize the sensitivity at $\mu=\omega_c\pm\omega_r$ and to maximize sensitivity at $\mu=\omega_c$.

\section{Data Analysis}\label{sec:data_analysis}
A set of data from one day is divided into groups, which typically each span several hours of data taking. For each group, we recompensate the flat surface and record CMOS images for an `X' shaped pattern [see Fig.~\hyperref[fig:flattening]{\ref*{fig:flattening}(a)}]. We also record CMOS images for all patterns imprinted in that group of data. The centering optimization described in Sec.~\ref{sec:CMOS} is performed for each group, and phase and amplitude maps for all patterns are calculated. The calculation of the phase maps only determines the phase modulo $2\pi$, which can result in artifacts as a full $2\pi$ phase shift is approached. We use the Scikit restoration.unwrap\_phase function to obtain a phase map without discrete jumps from $\pi$ to $-\pi$ or vice versa. 

Each group of data consists of multiple scans where we repeat the pattern imprinting sequence shown in Fig. 1(b) of the main text 250 times. The remainder of the analysis for each scan consists of the following steps with more detail provided below: (i) detect crystal reconfigurations and divide scan into segments, (ii) find the positions and sizes of the ions, (iii) determine the state of each ion in each shot and average over all shots in a segment, (iv) calculate the expected bright fraction using the phase and amplitude maps calculated using the CMOS images, (v) optimize the ACSS magnitude, size scale, relative beam phase for rotationally symmetric ($m=0$) patterns, and angle for non-rotationally symmetric ($m\neq0$) patterns, and (vi) calculate the difference between the measured bright state fraction and the prediction from the phase and amplitude maps generated with the CMOS images. After all of the data from the day are analyzed as above, the average of the optimal size scale factors for all patterns are taken, and steps (v) and (vi) are repeated, except with the size scale fixed to the day's average.

\begin{figure*}[t]
    \centering
    \includegraphics[width=.7\linewidth]{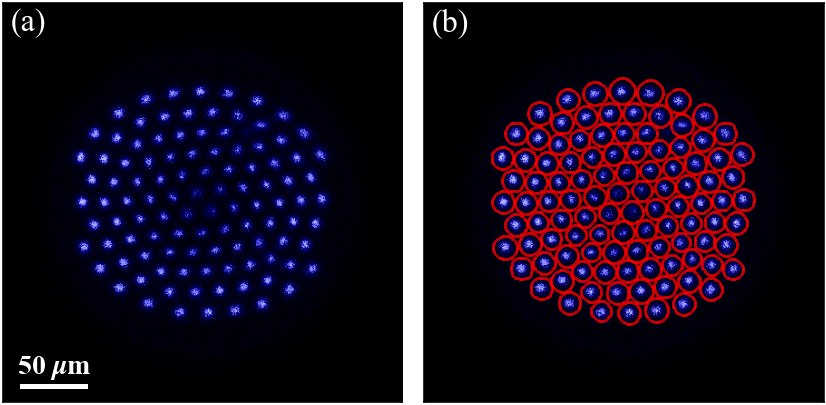}
    \caption{(a) Image of an ion crystal in the rotating frame. (b) The same crystal with ion regions marked. The center of a circle corresponds to the position found for the encircled ion.}
    \label{fig:ions}
\end{figure*}

\textbf{Steps (i)-(iii)--ion image analysis}: As described in the main text, we obtain images in the rotating frame of the crystal using a specialized camera \cite{Roentdek} that records photon arrival times in addition to the spatial coordinates \cite{britton2012,bohnet2016,wolf2024,mcmahon2024,jee2026,mitchell2001}.

Since the crystal rotation is well-controlled by the rotating wall \cite{huang1998}, we can trigger the time when an image starts recording on the signal used to control the rotation, so the crystal will always be in the same orientation in the image. We use the rotation frequency, photon arrival times, and photon spatial coordinates on the camera to obtain images in the rotating frame with clearly resolved ions.

During an experimental scan, we record images of the crystal while cooling. We compute the mean squared error between images from adjacent sets of 5 experimental repetitions to detect crystal reconfigurations. If a pair of images have a mean squared error above a threshold that is set before running the analysis code, that point in the scan is considered a reconfiguration. The segment with the largest number of repetitions between reconfigurations is then used for the analysis. The images recorded during Doppler cooling for the longest segment are added together. The resulting total Doppler cooling image is used for determining the position and regions of the ions [see Fig.~\ref{fig:ions}]. Further details for this computation are provided in Ref.~\onlinecite{bullock_thesis}.

As discussed in the main text, we perform a two-part detection sequence after each repetition of the experiment, where we record ion fluorescence for 25~ms, use the microwaves to flip the spins, and then record a second ion image for 25~ms. The counts in the second image in each ion's region are subtracted from those in the first image. A positive value corresponds to a bright state, while a negative value indicates a dark state. We use this protocol to mitigate the impact of the varying quantum efficiency across the active area of the camera. 

\begin{figure}
    \centering
    \includegraphics[width=\linewidth]{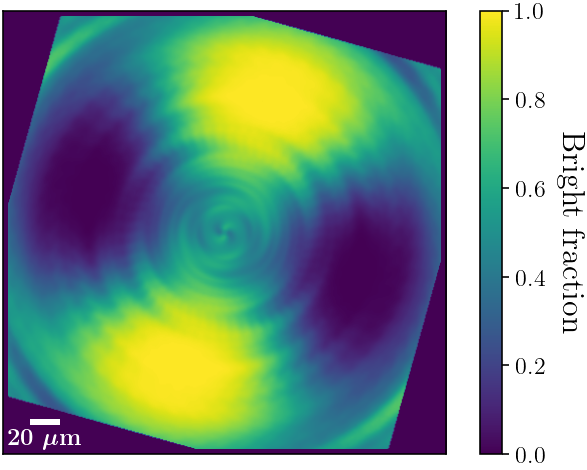}
    \caption{Predicted bright fraction for $Z_2^2,\,\varphi_{max}=2\pi$, \\$t=2t_{\pi/4}$.}
    \label{fig:bright_frac}
\end{figure}

\textbf{Step (iv)--imprinted pattern calculation}: We calculate the bright state fraction expected from the CMOS camera amplitude and phase maps by numerically integrating the spin precession based on the local phase and amplitude \cite{bullock_thesis},
\begin{align}
    \theta(\rho,\phi)&=\int_0^t\left\{ 2A\left(\rho,\phi-\omega_rt'\right)\right.\nonumber\\
    &\left.\times\sin\left[\Phi\left(\rho,\phi-\omega_r t'\right)-m\omega_r t'+\psi\right]dt'\right\}\label{eq:spin_precess_int}
\end{align}
Here $\theta$ is the integrated spin precession over the time interval $[0,t]$, $\rho$ and $\phi$ are radial and azimuthal coordinates in the rotating frame, and $A$ and $\Phi$ are the amplitude and phase maps. Equation~\ref{eq:spin_precess_int} is a generalization of Eq. 1 in the main text that accounts for amplitude inhomogeneitites. This is then mapped to a bright fraction $P_\uparrow(\rho,\phi)\approx\frac{1}{2}\left[1+\sin\left(2\theta\right)\right]$. We refer to the resulting bright fraction pattern as the predicted pattern. We also note that this calculation does not rely on the rotating wave approximation used to obtain Eq. 2 in the main text. More details of this computation are provided in Ref.~\onlinecite{bullock_thesis}. In Fig.~\ref{fig:bright_frac}, we show a predicted pattern calculated using the $Z_2^2,\,\varphi_{max}=2\pi$ amplitude and phase maps shown in Fig.~\ref{fig:pm_am} with an arm time of $t=2t_{\pi/4}$.

\begin{figure*}
    \centering
    \includegraphics[width=\textwidth]{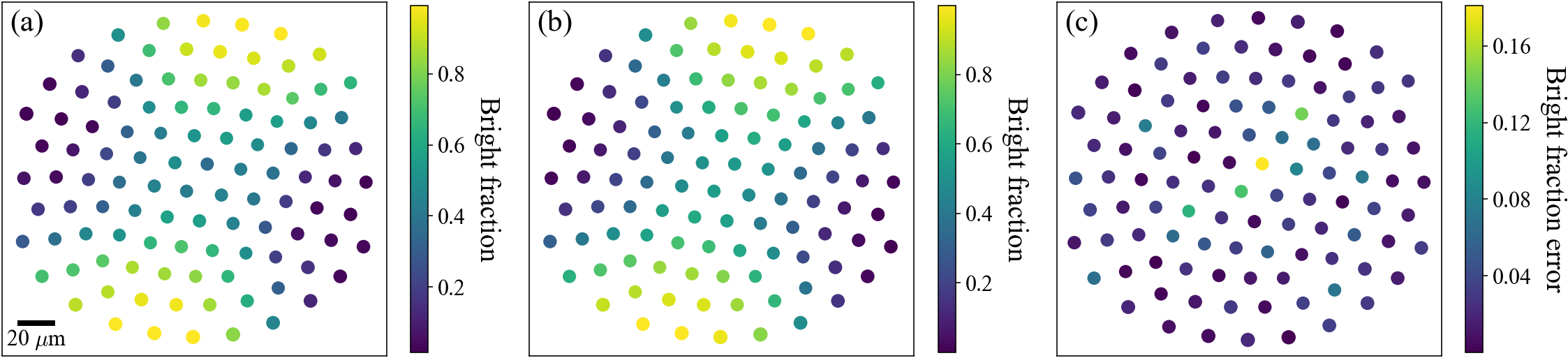}
    \caption{(a) Ion bright state fractions obtained from analyzing experimental images for the $Z_2^2$ pattern with $\varphi_{max}=2\pi$ and $t=2t_{\pi/4}$ used for Fig.~\ref{fig:ions} and Fig.~\ref{fig:bright_frac}. The circle sizes are proportional to, but not the same as, the ion regions shown in Fig.~\hyperref[fig:ions]{\ref*{fig:ions}(b)}. (b) Prediction of average bright fractions at each of the ions locations using the phase and amplitude maps obtained using the CMOS camera. (c) The absolute value of the difference between (a) and (b), indicating the error of each individual ion for this scan. This figure is the same as Fig. 3 in the main text.}
    \label{fig:ion_bright_fracs}
\end{figure*}

\textbf{Steps (v) and (vi)--error optimization and calculation}: The steps for determining the difference between an imprinted pattern and the corresponding predicted pattern, which we call the error, consist of taking an array describing the bright fraction of the ions [e.g. Fig.~\ref{fig:bright_frac}], the ion positions and regions for the ion crystal [Fig.~\hyperref[fig:ions]{\ref*{fig:ions}(b)}], and for each ion region, calculating the average predicted bright fraction in that area. This prediction is then compared to the measured bright fraction for the corresponding ion. If the error for each ion is $\epsilon_i$, the average error is calculated as the standard error, $\epsilon_{avg}=\frac{1}{\sqrt{N}}\sqrt{\sum_i\epsilon_i^2}$. In Fig.~\ref{fig:ion_bright_fracs}, we plot the measured and predicted bright fractions and the bright fraction errors at the location of each ion for the same $Z_2^2$ pattern shown in Fig. 2(c) of the main text. The amplitude and phase maps for this pattern were also shown in Fig.~\ref{fig:pm_am}, and the predicted pattern is shown in Fig.~\ref{fig:bright_frac}. Fig.~\hyperref[fig:ion_bright_fracs]{\ref*{fig:ion_bright_fracs}(a)} shows the average bright fraction obtained from the ion image analysis, and Fig.~\hyperref[fig:ion_bright_fracs]{\ref*{fig:ion_bright_fracs}(b)} shows the predicted bright fraction for each ion. Finally, Fig.~\hyperref[fig:ion_bright_fracs]{\ref*{fig:ion_bright_fracs}(c)} shows the error in the bright fraction of each ion, defined here as the absolute value of the difference between the values in shown in (a) and (b).

There are several parameters that affect the imprinted pattern, including the relative size scale of the ion camera coordinates and the CMOS camera coordinates, the ACSS amplitude $U$, and either the relative phase $\psi$ if $m=0$ or the angle of the imprinted pattern if $m\neq0$. While we estimate or calibrate these parameters separately, there can be inaccuracies in these starting values, so we optimize them to minimize $\epsilon_{avg}$. The size scale optimization is not performed when imprinting a uniform pattern, as there would be insufficient data for an accurate determination. The ACSS optimization is also performed differently for uniform and linear radial gradient patterns. For uniform patterns, the size scale is set to 1, and the ACSS scale is set to either 1 or 1.2 to account for the underestimation of $U$ for long arm times, based on which factor gives the lower error. The ACSS scale for radial gradients is determined by fitting the ion bright fraction as a function of the radius of ion $i$ $\rho_i$ to the analytic expression $P_\uparrow\left(\rho_i\right)=(1/2)\left\{1+\sin\left[2Ut\sin\left(k_\rho\rho_i+\psi \right)\right]\right\}$, where $k_\rho=2\varphi_{max}/\left(\rho_{max}\Lambda'\right)$, and $\Lambda'$ is the DM actuator pitch $\Lambda$ scaled by the demagnification factor of the telescope used to image the DM surface onto the ions, so approximately $\Lambda/10\sim30~\mu$m. This expression for $P_\uparrow$ can be straightforwardly calculated from the Hamiltonian for $m=0$ patterns, which is given in Eq. 3 in the main text.

The size scale estimate is obtained from an independent estimate of the imaging system magnification by comparing a representative crystal configuration with a simulation that determines the ion positions. From this, we estimate a magnification for the ion imaging system of 143. To account for potential uncertainty in this estimate and variations from the imperfect replication of the imaging telescope in the magnet bore vs on the optics table, during the initial analysis for a day we minimize the error between the calculated and measured bright fractions by adjusting the size scale. For a large majority of patterns, this correction is at most a few percent [see Sec.~\ref{sec:results_summ}]. When the data are reanalyzed using a fixed scale that is the average of all scans from that group of data, the average of $\epsilon_{avg}$ for all scans increases by about 0.01.

The ACSS amplitude is estimated based on a measurement of the time it takes to drive the first $\pi/2$ rotation after the beams are switched on. This time is divided by 2 to obtain the $\pi/4$ time $t_{\pi/4}$. However, we observe that the average time to drive subsequent $\pi/2$ rotations is shorter, so we expect that our ACSS estimate for patterns that are applied for longer arm times will be too small. Additionally, the intensity of the laser beams, which we do not actively stabilize, can fluctuate, again resulting in changes to the ACSS amplitude $U$, which in turn modifies the imprinted pattern. Similarly, if the intensity changes between when we imprint the pattern and when we record the images on the CMOS camera, the amplitude map $A$ would also change proportionally to the change in $U$. For each non-uniform pattern, we therefore calculate the imprinted pattern with a scaling factor for $U$ for a range of scaling factors. For shorter arm times, this range is 0.6 to 1.5, and for longer arm times, the range is 0.6-1.9. For many patterns, the dependence of the error on the ACSS scale is weak. If we find an optimal ACSS scale factor that results in a change in $\epsilon_{avg}$ of $<$0.01, we set the ACSS scale factor to 1. The variability in this parameter could be significantly improved by more frequent measurements of $t_{\pi/4}$, more careful measurements of the average $t_{\pi/4}$ for longer arm times, increasing arm times while reducing the ACSS to mitigate the impact of transient intensity changes when turning on the AOMs, and active intensity stabilization, which we already implement for most of our other beams. 

\begin{figure*}
\includegraphics[width=\textwidth]{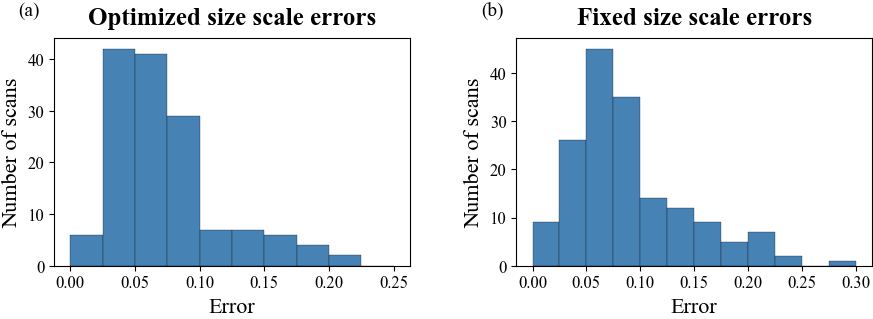}
\caption{Distribution of the differences between the measured and predicted qubit populations, $\epsilon_{avg}$. (a) Errors for all 144 scans for which the size scale was individually optimized. (This excludes patterns with the DM set to generate flat wavefronts.) (b) Errors for all 165 scans using fixed size scales (see text for details).}
\label{fig:size_scale}
\end{figure*}

The sensitivity to intensity fluctuations in our current configuration is exacerbated by the fact that each beam individually induces a uniform ACSS on the ions, $\Delta_z=\Delta_z^{(\textrm{DM})}+\Delta_z^{(\textrm{flat})}$, where $\Delta_z^{(\textrm{DM})} (\Delta_z^{(\textrm{flat}})$ are the ACSSs from the DM (flat) beam. If the DM (flat) beam has electric field amplitude $E_{(\textrm{DM})}$ ($E_{(\textrm{flat})}$), $\Delta_z^{b}\propto E_b^2\propto I_b$ for $b=$(DM) or (flat) and where $I_b$ is the intensity of a beam $b$. We can instead use a polarization gradient where the individual ACSS of each beam is nulled but there is still an effective ACSS when the beams are applied together \cite{britton2012} to reduce this sensitivity. For beams parallel to the magnetic field, if we write the polarization vector $\vec{\mathcal{E}}=a \hat{\sigma}^++b\hat{\sigma}^-$, we can find values $a$ and $b$ that null the ACSS of each beam individually at certain laser frequencies. However, if we set $a_{DM}=a,\,b_{DM}=b$ for the DM beam and $a_{flat}=a,\,b_{flat}=-b$ for the flat beam, the interference pattern between the beams still results in a nonzero ACSS. Currently we use $a=0,~b=1$ for both beams to maximize $U$ for a given laser power to reduce background light scatter off of the vacuum window.

To optimize the phase for the calculation of the predicted bright fraction for $m=0$ patterns, as discussed above, we calibrate the relative phase $\psi$ before starting an experiment by measuring the ACSS while scanning $\psi$ from 0 to $2\pi$ and track $\psi$ during the scans [see Sec.~\ref{sec:ion_calibrations}]. This procedure generally works well, such that the optimized phase is close to the set phase. However, for some patterns, potentially because of the phase map unwrapping procedure, the nominal phase is clearly incorrect, so we optimize this parameter as well. 
Finally, the angle of patterns with $m\neq0$ can vary. We therefore calculate the error for a variety of rotation angles and determine an optimal value.

\section{Data analysis summary}\label{sec:results_summ}
The data taken for this manuscript consist of 178 scans, 13 of which were excluded from the analysis results. Of these, 6 were removed due to hot ions or reconfigurations and 7 were removed due to technical issues with the synchronization of the rotating frame imaging or saving the images. Any scan where the large majority of the ions were clearly resolved was included, even if there were some imperfections in the ion localization. In Fig.~\hyperref[fig:size_scale]{\ref*{fig:size_scale}(a)}, we show a histogram of the errors from 144 of the remaining 165 scans with individually optimized size scales. Here uniform patterns are excluded because of insufficient data for an accurate size scale determination. In Fig.~\hyperref[fig:size_scale]{\ref*{fig:size_scale}(b)}, we show a historgam of errors for all 165 scans with fixed size scales determined by the average of the optimized size scales for a group of scans. When the scale is optimized for each scan [Fig.~\hyperref[fig:size_scale]{\ref*{fig:size_scale}(a)}], $82\%$ of scans have errors $<0.1$, while with fixed sized scales [Fig.~\hyperref[fig:size_scale]{\ref*{fig:size_scale}(b)}], $70\%$ of scans have errors $<0.1$. $7\%$ of the scans in Fig.~\hyperref[fig:size_scale]{\ref*{fig:size_scale}(b)} are uniform patterns with arm times $>45~\mu$s and errors $>0.1$ as expected due to the longer arm times (see Sec.~\ref{sec:error_sources}). This leaves about a $5\%$ residual difference in the percentage of scans with optimized vs fixed size scales that have errors $<0.1$.  Overall, the impact of the size scale optimization is small.

\begin{figure*}[t]
    \centering
    \includegraphics[width=\textwidth]{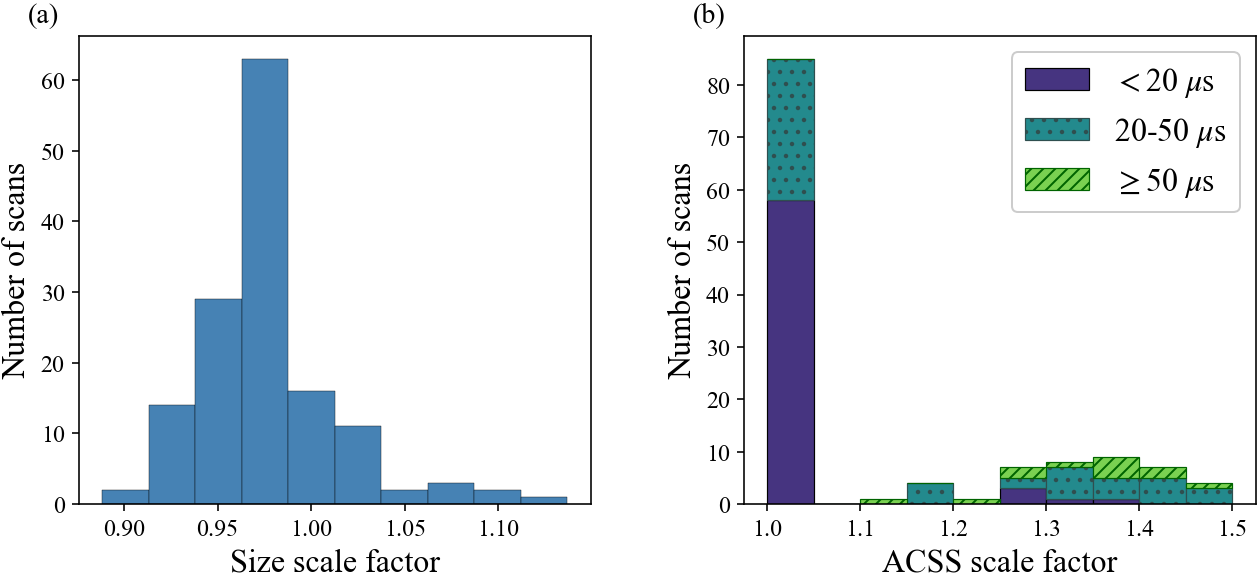}
    \caption{(a) Distribution of size scaling factors obtained with analysis of all scans except uniform patterns. (b) Distribution of ACSS scaling factors with arm times $<$20~$\mu$s ($<$$2t_{\pi/4}$) (purple solid), 20-50~$\mu$s ($\sim2t_{\pi/4}$-$4t_{\pi/4}$) (teal with dots), and $\geq$50~$\mu$s ($\geq$$4t_{\pi/4}$) (green with diagonal lines). Uniform and linear radial gradient patterns are excluded.}
    \label{fig:scaling_opts}
\end{figure*}

To look for trends, we plot the distribution of size and ACSS scale factors in Fig.~\hyperref[fig:scaling_opts]{\ref*{fig:scaling_opts}(a)} and Fig.~\hyperref[fig:scaling_opts]{\ref*{fig:scaling_opts}(b)}, respectively. Because of the different optimization approaches discussed in Sec.~\ref{sec:data_analysis}, we exclude uniform patterns from Fig.~\hyperref[fig:scaling_opts]{\ref*{fig:scaling_opts}(a)} and uniform and linear radial gradient patterns from Fig.~\hyperref[fig:scaling_opts]{\ref*{fig:scaling_opts}(b)}. The size scale factors for most scans were within a few percent of 1. Unlike the distribution for the size scale factors, the distribution for the ACSS scale factors is not centered around 1. However, of the scans included, about 70$\%$ have an optimal scale of 1, likely due to the fact that we set the scale to 1 if a different value does not improve the error by more than 0.01. Furthermore, we divide this histogram based on arm times $t$ to illustrate that the higher ACSS scale factors are associated with longer arm times. $\sim$90$\%$ of scans with $t<$20~$\mu$s, $\sim$50$\%$ of scans with 20~$\mu$s$\leq t<$50~$\mu$s, and 0$\%$ of scans with $t\geq$50~$\mu$s have ACSS scale factors of 1. It is clear that our current calibration is more likely to underestimate than to overestimate the ACSS amplitude, which we expect in part because of the short arm times for measuring $t_{\pi/4}$.

\begin{figure}
    \centering
    \includegraphics[width=\linewidth]{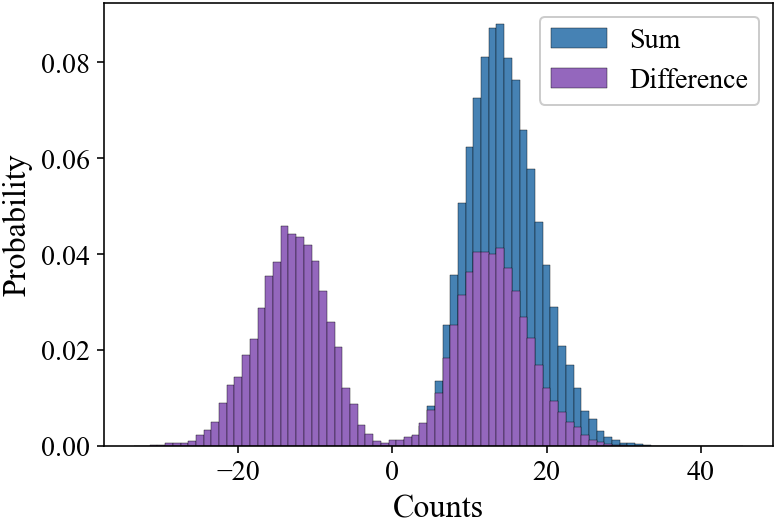}
    \caption{Histogram of detection counts for all ions in all images for an example scan where 105 images were analyzed. The blue histogram shows $c_{1}+c_{2}$, while the purple histogram shows $c_{1}-c_{2}$. The histograms are both normalized to a total probability of 1.}
    \label{fig:detect_hist}
\end{figure}

\section{Error sources}\label{sec:error_sources}
In this section we discuss several contributions to the discrepancy between the measured and predicted ion qubit populations.
\begin{figure*}
   \centering
   \includegraphics[width=\textwidth]{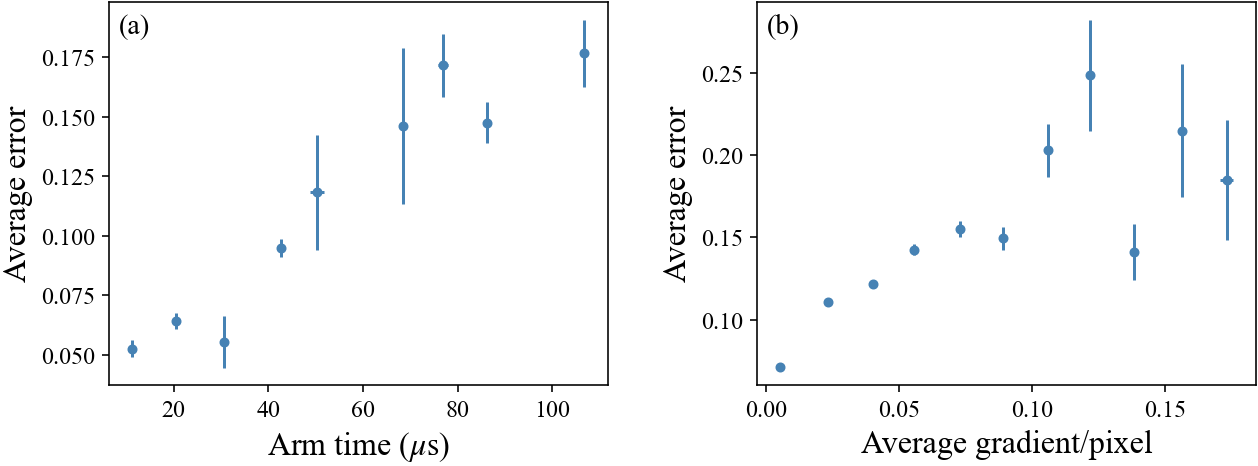}
   \caption{(a) Average pattern error vs arm time. Each point corresponds to the average of the error of all scans within a time bin of 12~$\mu$s, chosen to be close to a typical $t_{\pi/4}$. Both the horizontal and vertical locations are averages. Error bars in both axes are the standard error of the mean. (b) Average ion error vs the norm of the gradient of the bright fraction at the location of each ion. This does not account for the finite spatial size of the ions. Each point is binned in increments of 0.02. The pixel size is that of the ion camera--100~$\mu$m at the camera, which corresponds to about 0.7~$\mu$m at the ions. Error bars are the standard error of the mean.}
   \label{fig:armt_grad}
\end{figure*}
\textit{State readout errors}--As previously discussed, our detection sequence consists of two photon collection periods of 25~ms each with a microwave $\pi$ pulse in the middle. We define the counts recorded for each ion $i$ in the first detection interval as $c_{i_1}$ and those for the second interval as $c_{i_2}$. The average counts/ion in one 25~ms interval is $\sim$15. This number is limited by the quantum efficiency of our camera, which we estimate to be $<$10$\%$. In Fig.~\ref{fig:detect_hist}, we show a histogram of the detection results for all of the ions in one scan with a pattern where $\sim$50$\%$ of the ions are expected to be bright. The plot shows a histogram for $c_{i_1}+c_{i_2}$ in blue and one for $c_{i_1}-c_{i_2}$ in purple. The two peaks in the purple histogram correspond to bright ions [$c_{i_1}-c_{i_2}>$0] and dark ions [$c_{i_1}-c_{i_2}<0$] in each shot of the experiment. There are a small number ($<$1$\%$) of data points where the difference is equal to 0, which can occur, for example, if an ion is off-resonantly pumped from dark to bright between the first and second images. The rate of this process is determined by the detuning of the detection laser beams from any transitions from $\ket{\downarrow}$ to the $P_{3/2}$ manifold, which is tens of GHz for all transitions. The average counts in the blue histogram should be close to the average of the positive portion of the purple histogram. The small difference can be explained by an average of about 1 background count per shot per ion. For this particular scan, where there were 105 shots analyzed and 132 ions, $\left|c_{i_1}-c_{i_2}\right|<3$ for $\sim$1$\%$ of occurrences, indicating a very small error due to the overlap between the distributions for the bright and dark states.

To estimate the impact of projection noise, we consider the analysis of 100 images as an example. For ions that have a bright fraction of 0.5, the standard error of the mean of the measurement of the ion's spin state due to projection noise would be about 0.06, which constitutes a significant error. This number would be significantly improved by the ability to take more images between reconfigurations, which, if we hold the desired number of photons per ion per image constant, could be achieved with a camera with a higher quantum efficiency.

\textit{Spin decoherence and spontaneous emission}--We independently measure a spin-echoed spin decoherence due to magnetic field noise of $e^{-\left(\Gamma_{\textrm{spin}}t\right)^2/2}$ where $t$ is the arm time of the spin-echo sequence used for the measurement and $\Gamma_{\textrm{spin}}\sim$100~s$^{-1}$. To measure the decoherence due to off-resonant light scattering, we apply either the DM beam (with the DM surface set to flat) or flat beam for varying arm times and fit to a decaying exponential that accounts for both the spin decoherence and the spontaneous emission, $e^{-\left(\Gamma_{\textrm{spin}}t\right)^2/2}e^{-\Gamma_{\textrm{emis}}t}$, where $\Gamma_{\textrm{spin}}$ and $\Gamma_{\textrm{emis}}$ are the decay rates for the spin decoherence and spontaneous emission, respectively, and $t$ is the arm time. A typical decay rate due to spontaneous emission for an individual beam with an approximately 7~kHz ACSS is $\Gamma_{\textrm{emis}}\sim$140~s$^{-1}$. The decay rate resulting from off-resonant scattering is doubled when both beams are applied. A calculation of the decoherence using this expression and these values gives a $0.003$ error for a 10~$\mu$s arm time and a $0.035$ error for a 125~$\mu$s arm time, which is longer than any arm time used for imprinting patterns. Because the intensity of the beams is lower for the first few microseconds as evidenced by the underestimation of the ACSS for short arm times discussed above, this calculation likely overestimates the error for short arm times, so it should be negligible for results where $t=t_{\pi/4}\sim$11~$\mu$s.

\begin{figure*}
    \centering
    \includegraphics[width=\textwidth]{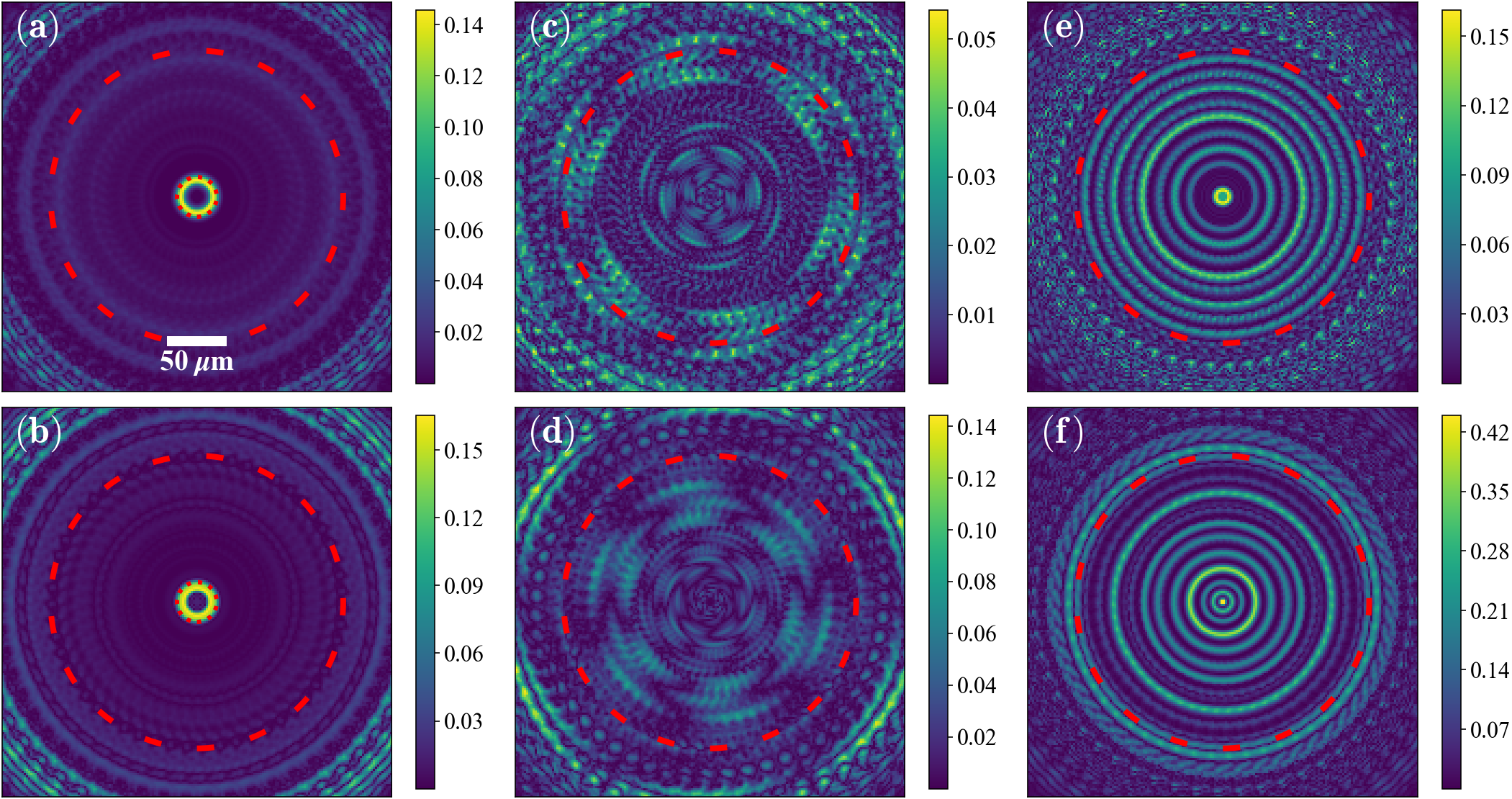}
    \caption{Calculated bright fraction differences due to misalignment through the $f=95$~mm used to image the DM surface onto the CMOS camera. The top row corresponds to a 3~mm vertical displacement (perpendicular to the beam path) of the lens, while the bottom row corresponds to a 0.5~mm misalignment of the focus. All calculations assume $t=t_{\pi/4}$. Large red dashed circles indicate a radius of 125~$\mu$m, an upper limit on the crystal radius. (a) and (b) Ring, $R=0,\,\varphi_{max}=\pi$ [Fig. 2(a) in the main text]. The small red dashed circles indicate the approximate position of the first ring of ions around the center, which experiences a nonzero ACSS. (c) and (d) $Z_5^3,\,\varphi_{max}=2\pi$ [Fig. 2(d) in the main text], (e) and (f) Radial gradient, $2\varphi_{max}/\rho_{max}=2.6\pi,\,\rho_{max}=5$.}
    \label{fig:lens_centering}
\end{figure*}

\textit{Arm time and pattern gradient errors}--In Fig.~\hyperref[fig:armt_grad]{\ref*{fig:armt_grad}(a)}, we plot the average pattern error vs arm time, and see a clear trend. The decoherence due to spontaneous emission and spin decoherence is insufficient to account for this change. In general, the sharpness of the features in the pattern is also correlated with increased errors, and longer arm times will result in sharper features for a given pattern. Furthermore, the errors from misalignment discussed below will have a larger effect for longer pattern application times. 

In Fig~\hyperref[fig:armt_grad]{\ref*{fig:armt_grad}(b)}, we plot the average norm of the gradient of the predicted bright fraction at the location of each ion in each scan. An increase in errors with steeper gradients is expected, since the motion and exact position of the ion will have a larger impact on the effective ACSS. The statistics for the larger gradients preclude any strong conclusions, but, as intuition would suggest, there appears to be a correspondence between the local steepness of the pattern and the error or discrepancy between the predicted and measured populations of an ion. One potential improvement to the errors due to gradients would be improving the accuracy with which we determine the ion locations. Small discrepancies in the position of ions located in a steep ACSS gradient will have larger impacts on the errors, whereas the exact position is less critical for predicting the bright fraction for an ion at a location with a small gradient. These data also demonstrate the necessity of good cooling of the in-plane motion of the ions for this technique to reach its maximum potential, since colder ions are better localized.

\textit{Lens alignment}--Because of the constraints imposed by the location of the lens that is located in the bore of the superconducting magnet, we have limited adjustment capabilities and cannot directly observe the lateral alignment of this lens in the beams. On the other hand, we try to center the beams going through the same lens on the table in front of the CMOS camera. We can evaluate the impact of potential misalignment through the in-bore lens by centering the beam on the lens on the table, translating the lens by known amounts, and comparing the resulting predicted bright fraction patterns. We begin with the beam well-centered on the lens. In this position, we first ensure the DM surface is set to compensate any deviations from flat wavefronts using the procedure discussed in Sec.~\ref{sec:CMOS}. We then record images while varying the relative phase between the DM and flat beams for a variety of patterns. These images are used to generate phase and amplitude maps. We repeat this procedure at several lens positions, including displacements both transverse and parallel to the beam path. For each position and pattern, we compute phase and amplitude maps after redetermining the center pixel. We assume an arm time $t=t_{\pi/4}\sim11~\mu$s, set other relevant parameters to typical values, and calculate the expected bright fraction using Eq.~\ref{eq:spin_precess_int}. We then evaluate the impact of the lens displacements by subtracting the expected bright fraction map at the centered lens position from the bright fraction map with the translated lens. 

\begin{figure*}
\centering
\includegraphics[width=\textwidth]{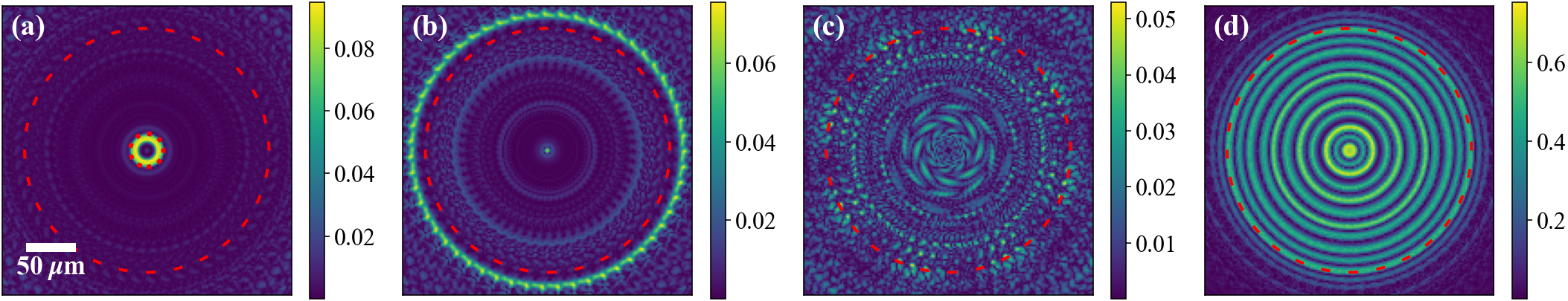}
\caption{Calculated bright fraction differences when the ACSS pattern center is displaced by two CMOS camera pixels, equivalent to a misalignment of $3.7~\mu$m, for an arm time of $t_{\pi/4}$, representing the pattern being off center from the rotation axis of the ion crystal. Large, red, dashed circles indicate a radius of 125~$\mu$m, an upper limit on the crystal radius. (a) Ring, $R=0,\,\varphi_{max}=\pi$ [Fig. 2(a) in the main text]. The small, red, dashed circle indicates the approximate position of the first ring of ions around the center, which experience a nonzero ACSS. (b) Ring, $R=4,\,\varphi_{max}=\pi$ [Fig. 2(b) in the main text]. (c) $Z_4^4,\,\varphi_{max}=2\pi$. (d) Radial gradient, $2\varphi_{max}/\rho_{max}=2.6\pi,\,\rho_{max}=5$.}
\label{fig:ion_centering}
\end{figure*}

Figure~\ref{fig:lens_centering} shows a selection of results of this analysis. The top row shows the impacts of displacing the lens 3~mm transverse to the beam direction, while the bottom row shows the results from a 0.5~mm longitudinal lens displacement. The patterns shown include a ring with $R=0$ and $\varphi_{max}=\pi$ [ion pattern shown in Fig. 2(a) in the main text] [(a) and (b)], $Z_5^3,\,\varphi_{max}=2\pi$ [ion pattern shown in Fig. 2(d) in the main text] [(c) and (d)], and a linear radial gradient with $2\varphi_{max}/\rho_{max}=2.6\pi$ and $\rho_{max}=5$ [(e) and (f)]. The larger red circles indicate a radius of 125~$\mu$m to illustrate an upper limit on the maximum radial extent of the crystals. The smaller red circle in Fig.~\hyperref[fig:lens_centering]{\ref*{fig:lens_centering}(a)} corresponds to an approximate radius for the ring of ions immediately around the center ion, which, as is shown in Fig. 2(a) in the main text, experience a nonzero ACSS with this ring pattern. Both the transverse and longitudinal displacements have similar impacts on the ring, with maximum differences of around 0.15 within a small region. For the Zernike polynomial, the vertical displacement only causes differences of $\lesssim0.05$. The difference in the Zernike polynomial with the misalignment of the focus is similar to the maximum differences in the ring pattern, but the most severe impacts are far enough from the center of the pattern (well outside the red dashed circle) that they would have minimal effects on the ions. On the other hand, while the radial gradient has a similar maximum error for the vertical displacement, the differences are large in the center as well as near the edges, so the misalignment will have a more significant effect. Moreover, the focal misalignment has an extreme impact on the radial gradient, with discrepancies in the center of $\sim$0.4.

\textit{Centering on ions}--We expect that the technique for aligning the central actuator of the DM with the rotation axis of the ion crystal described in Sec.~\ref{sec:ion_calibrations} results in an alignment within $\sim$1$-$2~$\mu$m, but here we examine the impact of larger misalignments to illustrate the importance of the centering procedure. Again, we look at several patterns. For each pattern, we use a phase and amplitude map obtained from the CMOS images that have been centered using the procedure involving the `X'-shaped pattern described in Sec.~\ref{sec:data_analysis}. We calculate the predicted bright fraction with an arm time of $t=t_{\pi/4}\sim11~\mu$s for the phase and amplitude maps when they are well-centered and shift the calculated phase and amplitude maps by 2 CMOS pixels (1.85~$\mu$m per pixel) vertically in software and recalculate the predicted bright fractions. We then take the absolute value of the difference between the off-center and centered predicted bright fractions.

We show the resulting plots of the differences for 4 patterns in Fig.~\ref{fig:ion_centering}. In Fig.~\hyperref[fig:ion_centering]{\ref*{fig:ion_centering}(a)-(b)} we show the difference in the expected bright fraction for ring patterns with a maximum phase shift $\varphi_{max}=\pi$ and radii $R=0$ and $R=4$, respectively. Corresponding ion images for these two patterns are shown in Figs. 2(a) and 2(b) in the main text. We also show the bright fraction difference for $Z_4^4$ with $\varphi_{max}=2\pi$ in Fig.~\hyperref[fig:ion_centering]{\ref*{fig:ion_centering}(c)} and for a linear radial gradient with $2\varphi_{max}/\rho_{max}=2.6\pi,~\rho_{max}=5$ in Fig.~\hyperref[fig:ion_centering]{\ref*{fig:ion_centering}(d)}. As in Fig.~\ref{fig:lens_centering}, the large red circles represent an upper limit on the crystal radius, while the smaller circle in (a) shows an approximate radius for the first ring of ions around the center ion. For the first three patterns shown, there is a clear effect, but the difference from the ideal case remains $<0.1$ throughout the pattern. The radial gradient shown in Fig.~\hyperref[fig:ion_centering]{\ref*{fig:ion_centering}(d)}, however, is strongly impacted by being slightly off-center. Therefore, for steep radial gradient patterns, we would expect large errors to be introduced by only a small misalignment. The fact that we observe errors in radial gradient patterns of typically $<$0.2 adds support for our estimate of alignment accuracy better than about 2~$\mu$m. On the other hand, this analysis suggests that for many other patterns, this protocol is robust to displacements of several micrometers.

\end{document}